\documentclass[aps,twocolumn,prl,showpacs]{revtex4}
\usepackage{graphicx}
\usepackage{epsfig}
\usepackage{amsmath}
\usepackage{graphics}
\usepackage{latexsym}
\begin{document}

\title{Anisotropic interfacial tension, contact angles, and line tensions: A graphics-processing-unit-based Monte Carlo study of the Ising model}

\author{Benjamin J. Block, Suam Kim, Peter Virnau and Kurt Binder}

\affiliation{Institut f\"ur Physik, Johannes Gutenberg-Universit\"at Mainz, Staudinger Weg 9, D-55099 Mainz, Germany}

\begin{abstract}
As a generic example for crystals where the crystal-fluid interface tension depends on the orientation of the interface relative to the crystal lattice axes, the nearest neighbor Ising model on the simple cubic lattice is studied over a wide temperature range, both above and below the roughening transition temperature. Using a thin film geometry $L_x \times L_y \times L_z$ with periodic boundary conditions along the z-axis and two free $L_x \times L_y$ surfaces at which opposing surface fields $\pm H_{1}$ act, under conditions of partial wetting, a single planar interface inclined under a contact angle $\theta < \pi/2$ relative to the yz-plane is stabilized. In the y-direction, a generalization of the antiperiodic boundary condition is used that maintains the translational invariance in y-direction despite the inhomogeneity of the magnetization distribution in this system. This geometry allows a simultaneous study of the angle-dependent interface tension, the contact angle, and the line tension (which depends on the contact angle, and on temperature). All these quantities are extracted from suitable thermodynamic integration procedures. In order to keep finite size effects as well as statistical errors small enough, rather large lattice sizes (of the order of 46 million sites) are found necessary, availability of very efficient code implementation of graphics processing units (GPUs) was crucial for the feasibility of this study.
\end{abstract}
\pacs{68.03.Cd, 75.10.Hk, 02.70.Uu}
\maketitle

\section{Introduction}
Heterogeneous nucleation of crystals from the vapor phase or from the melt is a physically important process in many circumstances, from the formation of ice on dust particles in the atmosphere \cite{1} to the processing of alloys in metallurgy \cite{2,3}. A key aspect here is the fact that the solid-vapor or solid-liquid interfacial tensions are anisotropic, i.e. the interfacial excess free energy depends on the orientation of the interface normal relative to the crystal axes. Under these circumstances, the standard concepts of homogeneous nucleation theory \cite{3,4,5,6,7,8} assuming a spherical shape of the nucleus of the stable phase, and of heterogeneous nucleation on a flat substrate assuming a sphere cap shape of the nucleus \cite{9,10,11,12} are no longer valid. Also Young's equation \cite{13} which expresses the equilibrium contact angle of a droplet at a substrate (wall) in terms of the wall excess free energies of the coexisting phases and their interfacial tension ceases to be valid \cite{14,15,16,17}.

A general solution for this problem of equilibrium crystal shape, of a crystal coexisting with surrounding bulk vapor or liquid phase has been provided by the Wulff construction \cite{18,19,20,21,22,23,24,25}. This solution has been extended to wall-attached crystallites by the Winterbottom construction \cite{26}. However, the Wulff construction needs the full knowledge of the orientation-dependent interfacial free energy $\gamma (\Omega)$, where $\Omega$ stands for the polar angles $\vartheta$, $\varphi$ relative to a suitable reference axis (e.g., $\vartheta§ = 0$ if the interface plane coincides with the close-packed 111 surface of a face-centered cubic crystal, and $\varphi =0$ if the x-axis in that plane coincides with a symmetry axis of the crystal). The Winterbottom construction needs the additional knowledge of the anisotropic wall excess free energies of the coexisting phases. Finally, we stress that this knowledge not only is needed for the understanding of equilibrium crystal shapes, but also for crystal growth phenomena \cite{27,28}, which may happen under far from equilibrium conditions, and there is a complicated interplay of these anisotropic excess free energies, which enter the driving forces for transport processes, with various transport coefficients.

Unfortunately, explicit predictions of such interfacial excess free energies of crystals at nonzero temperature are scarce and difficult to obtain. This lack of knowledge severely hampers the understanding of heterogeneous nucleation of solids as well as of the further growth of these nuclei and the resulting microstructure formation of materials \cite{29}.

In the present paper, we address this issue of the anisotropic interfacial tension and the associated contact angle that arises for such interfaces at planar walls for a simplistic but generic model system, namely the Ising model with nearest neighbor (ferromagnetic) interaction on the simple cubic lattice. This system is useful, since for the related two-dimensional problem (Ising square lattice) the problem of the anisotropic interfacial tensions has been solved exactly, and hence the shapes of the resulting two-dimensional crystalline domains are explicitly known \cite{23,24,25}.

However, this two-dimensional system is an exception also due to the fact that the crystal shapes contain strictly linear pieces only at zero temperature: i.e., at nonzero temperature no crystal facets can occur. This is no longer true in the three-dimensional Ising model, where facets occur at temperatures $T <T_R$, the critical temperature of the interface roughening transition \cite{30,31}. For $T \geq T_R$, no facets occur, and for $T \rightarrow T_c$, the critical temperature of the Ising model, anisotropy effects vanish, both in $d=2$ and in $d=3$ dimensions \cite{32}. Previous work on homogeneous and heterogeneous nucleation in the Ising model (e.g. \cite{33,34}) has ignored anisotropy effects throughout (with the notable exception of \cite{35}).

Since nucleation phenomena deal with nuclei of nanoscopic size, in the case of wall-attached nuclei the effect of the line tension of the three-phase contact line (where the interface between the coexisting phases meets the wall \cite{36,37,38,39,40,41,42}) must not be neglected. In the present paper, where we apply a Monte Carlo computer simulation approach \cite{43,44,45} and thus necessarily deal with finite systems and therefore need to address finite size effects \cite{43,44,45,46,47}), it is hence necessary to include the possibility of line tension effects in the analysis.

The outline of this paper is as follows. In Sec.~II, we describe the chosen simulation geometries and discuss the general methodology of extracting the various excess free energies (e.g., by thermodynamic integration \cite{45} methods, controlling the wall excess free energies by variation of boundary fields). Sec.~III describes our numerical results, emphasizing also the effects of interfacial anisotropy on essentially planar interfaces (for $T>T_R$ one must distinguish between interfacial tensions and interfacial stiffness \cite{48,49,50} which we characterize; for $T<T_R$ the free energy cost of interfacial steps (see e.g. \cite{51,52}) is computed.) Sec.~IV, for the first time, gives comprehensive results on the dependence of the line tension on both temperature and contact angle, while Sec.~V summarizes our conclusions. This study has required a massive investment of computer resources, which was possible only due to very efficient codes on graphics programming units (GPUs); the main features of our respective code development are summarized in an appendix.

\section{Model Geometry and Simulation Methodology}
\subsection{The need of a generalization of antiperiodic boundary conditions}

We consider Ising systems with nearest neighbor ferromagnetic exchange interaction $J$ on the simple cubic lattice, assuming a geometry of the simulation box with linear dimensions $L_x, L_y$ and $L_z$ in the three directions $(x,y,z)$ of the coordinate system (Fig.~1). We assume two free surfaces of area $L_y L_z$, on which antiparallel surface magnetic fields act, $H_1<0$ and $H_n=-H_1>0$. If we choose the lattice spacing as the unit of length, the index $n$ of the last layer in x-direction simply is $n=L_x$. If, for the sake of generality, we also allow for a field $H$ that acts on all the spins and hence breaks the symmetry, the Hamiltonian is

\begin{equation}\label{eq1}
\begin{split}
\mathcal{H} = - J \sum \limits _{\langle i,j\rangle} S_iS_j - H \sum \limits _i S_i - H_1 \sum \limits _{i \in 1} S_i - H_n \sum \limits _{i \in n} S_i,\\ 
\quad S_i = \pm 1 \;,
\end{split}
\end{equation}

and the notation $\langle i,j \rangle$ means that over all pairs of spins is summed once. Note that in the z-direction we use periodic boundary conditions (PBC) throughout, while in y-direction several choices of boundary conditions \cite{53} shall be considered.

The simplest choice is the choice of PBC in y-directions as well. This choice is the standard choice that has been widely implemented to study wetting phenomena \cite{54,55,56} and interface localization-delocalization transitions \cite{57,58,59} in the Ising model \cite{55} provided the thermodynamic limit is taken in the appropriate manner. In the present context, we need to consider this choice to obtain surface excess free energy differences that arise due to the action of the surface fields in the layers $k=1$ and $k=n$ in Eq.~\ref{eq1}. This choice is also appropriate for the study of contact angles, when the anisotropy of the interfacial free energy can be neglected, and therefore simply the Young equation \cite{13} for the contact angle can be used \cite{33,34}. Unlike the situation alluded to in Fig.~1, one then considers only states which (for the case of partial wetting \cite{57,58} are single phase ``spin up'' or ``spin down'' phases at bulk field $H=0$), apart from the region near the free surfaces where the surface fields $H_1,H_n$ act.

In order to see this, we note that for $L_x \rightarrow \infty$ the free energy per spin at temperature $T$ can be decomposed as

\begin{equation}\label{eq2}
\begin{split}
f(T,H,H_1,H_n,L_x)=f_b(T,H) + \frac {1}{L_x} f_1 (T,H,H_1) + \\
\frac {1}{L_x} f_n(T,H,H_n),
\end{split}
\end{equation}

where the bulk free energy per spin $f_b(T,H)$ is independent of the boundary fields, of course. The surface excess free energy due to the free surface at layer $n=1$ is denoted as $f_1(T,H,H_1)$ and can depend on the field $H_1$ but not on the field $H_n$. Likewise, the surface excess free energy due to the free surface at layer $n$, $f_n(T,H,H_n)$ does not depend on $H_1$. Of course, the strict separation of $f$ into a bulk term $f_b$ and two boundary corrections (small of order $1/L_x)$ holds only in the limit of very large $L_x$, when correlations between magnetization fluctuations in the layers $k=1$ and $k=n$ are negligibly small.

When we consider the limit $H \rightarrow 0^+$, the magnetization $m_b=-(\partial f_b(T,H)/\partial H)$ for temperatures $T$ less than the critical temperature $T_{cb}$ in the bulk is the positive spontaneous magnetization, $m_b= + m_{coex}$, while in the limit $H \rightarrow 0^-$ we obtain $m_b= -m_{coex}$. The direction of the double arrow in Fig.~1 symbolizes these two possible states in the bulk, and in Fig.~1 it is assumed that the interface separating the two domains is inclined under an angle $\theta$ relative to the (z,y) plane.

However, for the moment we wish to discuss boundary effects on the pure phases in the thin film (with PBC in the y-direction only an even number of interfaces separating domains in the bulk obviously is possible). Of course, in the bulk phase coexistence is only possible for $H=0$, and $f_b ^{(+)} (T,0) = f_b^{(-)}(T,0)$, when we distinguish the coexisting phases with magnetization $+m_{coex}$ by a superscript (+), and with magnetization $-m_{coex}$ by a superscript (-). However, the surface excess free energies $f_1(T,0,H_1)$ and $f_n(T,0,H_n)$ are well-defined also for nonzero surface fields, and we notice the symmetry relations

\begin{equation}\label{eq3}
f_1^{(+)} (T,0,H_1) = f_n^{(-)} (T,0,H_n=-H_1),
\end{equation}
\begin{equation}\label{eq4}
f_1^{(-)} (T,0,H_1) = f_n^{(+)} (T,0,H_n=-H_1).
\end{equation}

These symmetries reflect the symmetry of the Hamiltonian, Eq.~(1), against a simultaneous change of sign of all the spins (which transforms the superscript (+) into (-) and vice versa) together with the sign of the surface fields. The special choice $H_n=-H_1$ also shows up in a special symmetry of the magnetization profile $m(x)$ across the film, namely (cf. Fig.~2)

\begin{equation}\label{eq5}
m^{(+)} (x,H_1,H_n = -H_1) = - m ^{(-)} (L_x -x, H_1,H_n=-H_1).
\end{equation}

A consequence of this general symmetry relation yields for the surface layer magnetizations $m_1$, $m_n$ readily

\begin{equation}\label{eq6}
m_1^{(+)} (H_1) = -m_n^{(-)}(H_n) \quad , \quad m_1^{(-)} (H_1)=-m_n^{(+)} (H_n).
\end{equation}

Note that these surface layer magnetizations depend on the surface fields in their respective layer, of course, since we also have (the arguments $T,H$ are here suppressed for simplicity)

\begin{equation}\label{eq7}
m_1=-(\partial f_1/\partial H_1)_{T,H} \quad , \quad m_n=-(\partial f_n/\partial H_n)_{T,H}.
\end{equation}

At this point we note that for the Ising model Young's equation \cite{13} for the contact angle $\theta$ can be cast in the form \cite{33,34}

\begin{equation}\label{eq8}
f_{int}(T) \cos \theta = f_1^{(+)} (T,0,H_1) - f_1^{(-)} (T,0,-H_1) = \Delta f_1
\end{equation}

Eqs.~\ref{eq6}, \ref{eq7}, can be combined into a form useful for simulations, namely

\begin{equation}\label{eq9}
\Delta f_1=- \int \limits _0^{H_1} dH_1'[m_1(H_1')+m_n(H_1')]
\end{equation}

and hence the right hand side of Eq.~\ref{eq8} is readily obtained, and it is generally true that for $H_1=H_n=0$ we must have $\theta = \pi/2$ (the profiles in Fig.~2 then have the additional symmetry $m^{(+)}(x) = -m^-(x)$), and hence must also be symmetric with respect to the interchange of x and $L_x-x$, cf. Eq.~\ref{eq5}. However, in the case of an anisotropic interface tension $f_{int}(T,\theta)$ Young's equation is no longer true \cite{14,15,16,17}, as will be discussed below.

When we now wish to consider a state where we have two domains which have the magnetization profiles described in Eq.~\ref{eq5} and Fig.~2 separated by a single interface, as shown in Fig.~1, we need a suitable boundary condition in y-direction, which seamlessly transforms the up-domain on the left side into the down-domain on the right side. ``Seamlessly'' implies here that the symmetry relation written in Eq.~\ref{eq5} results, of course.

Writing the index i labelling the position of a spin on the lattice in terms of its Cartesian coordinates x,y,z we obtain

\begin{equation}\label{eq10}
S(x,y,z) = -S(L_x-x, y \pm L_y, z).
\end{equation}

Eqs.~\ref{eq5}, \ref{eq10}, using coordinates in the continuum imply a simulation box where $0 \leq x \leq L_z$, surface potentials acting at $x=0$, and $x=L_x$, respectively (e.g., this could be realized for a binary symmetric Lennard-Jones mixture, confined between ``antisymmetric'' walls, where one wall attracts A particles while the other wall attracts B-particles, and the spin variable $S(x,y,z) = \pm 1$ denotes whether the particle located at a point $(x,y,z)$ is of type A or of type B). Dealing with a lattice, we have discrete lattice planes $k_x=1,2,\ldots,L_x$ in x-direction, and then Eq.~\ref{eq10} is replaced by

\begin{equation}\label{eq11}
S(k_x,k_y,k_z) = -S(L_x+1-k_x, k_y \pm L_y, k_z)
\end{equation}

and the magnetization profile is (we now do not write the boundary fields explicitly)

\begin{equation}\label{eq12}
m^{(+)} (k_x) = -m^{(-)} (L_x+1-k_x) \quad , \quad k_x =1,\ldots,L_x
\end{equation}

with $m^{(+)} (1) = m_1^{(+)}, m^{(+)} (L_x)=m_n^{(+)}, m^{(-)} (1) = m_1^{(-)}, m^{(-)} (L_x)=m_n^{(-)}$ according to our previous notation \{Eq.~\ref{eq6}\}. We denote Eqs.~\ref{eq10}, \ref{eq11}, as generalized antiperiodic boundary condition (GAPBC). To our knowledge, this has not been implemented in the literature before; related twisted boundary conditions (without change of sign of the spin across the boundary) have been used to study Ising criticality at the Moebius strip and at the Klein bottle \cite{53}. We emphasize that with this boundary condition we create full translational invariance in y-direction, i.e., when the total magnetization of the system is not conserved, the interface in Fig.~1 could be arbitrarily translated in y-direction. However, when the magnetization is conserved, and required to be zero, then the average position of the interface is fixed to be in the middle of the box (at $x=L_x/2$ the interface is at $y=L_y/2$, irrespective of the choice of $H_1$, as long as the contact angle $\theta$ is large enough so that $L_x \cot \theta < L_y$, so the inclined interface fits into the system.) Of course, the geometry chosen is not suitable to study the wetting transition, where $\theta \rightarrow 0$ \cite{54,55,56} and the equilibrium situation of a thin film confined between walls with competing surface fields would be an interface oriented parallel to the confining walls \cite{55,57,58,59}.

\subsection{Interfacial free energy contributions in confined Ising films with generalized antiperiodic boundary conditions}

We now discuss the free energy of an Ising film in the geometry of Fig.~1. In comparison with a system with the same linear dimensions $L_x,L_y,L_z$ but with periodic boundary conditions throughout, in a state with uniform positive $(+)$ or negative (-) spontaneous magnetization, we encounter excess terms in the free energy due to the walls, the inclined interface, and the two contact lines (of length $L_z$) when the interface meets the wall. The interfacial area is $\tilde{L} L_z$, with $L_x= \tilde{L} \sin \theta$ (cf. Fig.~1). Note that for $H_1=0$ the interface would be perpendicularly oriented, $\theta = \pi/2$, and when we consider the case of conserved magnetization $m=0$, both the (+) domain and the (-) domain have two free surfaces of area $L_zL_y/2$ each. However, when the boundary field is nonzero, it is energetically more favorable to tilt the interface, such that the surface of the (+) domain where the negative surface field acts gets smaller, namely $L_z(L_y-\tilde{L} \cos \theta)/2$, and the surface of the (-) domain gets larger, $L_z(L_y + \tilde{L} \cos \theta)/2$, see Fig.~1. On the top wall, where the positive boundary field acts, the situation is opposite. Note that the antisymmetry of the boundary fields $(H_n=-H_1)$ ensures that the interface between the domains with opposite sign of the magnetization is inclined but planar, while for the case of symmetric fields $(H_n=H_1)$ a curvature of the interface would be expected, which would give rise to additional correction terms \cite{60,61,62,63}.

Of course, due to the special spin reversal symmetry of the Ising model this problem that domain walls across a thin film in general will exhibit curvature can be avoided, with the choice of the antisymmetric surface fields $H_n=-H_1$, as used here.

These considerations allow to write down the total interfacial and line excess free energies of the system as follows (remember $H_1 <0$; the interface tension of the tilted interface will depend on the angle $\theta$ and is denoted as $\gamma (\theta)$)

\begin{equation}\label{eq13}
\begin{split}
F_{int} = \tilde{L} L_z \gamma (\theta) + \frac 1 2 L_z (L_y + \tilde{L} \cos \theta) \gamma ^{(+)} (|H_1|)  + \\
\frac 1 2 L_z(L_y-\tilde{L} \cos \theta) \gamma ^{(-)} (|H_1|) +\\
\frac 1 2 L_z (L_y - \tilde{L} \cos \theta) \gamma ^{(+)}(-|H_1|) +\\
 \frac 1 2 L_z (L_y + \tilde{L} \cos \theta) \gamma ^{(-)} (-|H_1|)+
2 L_z \tau (|H_1|,\theta)
\end{split}
\end{equation}

In the last term, we have defined the line tension $\tau$ (which may depend on $|H_1|$ and $\theta$).

Here we have also simplified the notation by redefîning $f_1^{(+)} (T,0,H_1) = f_n^{(+)}(T,0,H_1) = \gamma ^{(+)} (H_1), \; f_1^{(-)} (T,0,H_1)=f_n^{(-)}(T,0,H_1)= \gamma ^{(-)}(H_1)$, noting that the two surfaces at $k_x=1$ and $k_x=n$ are physically equivalent. Utilizing then the symmetries noted in Eqs.\ref{eq3}, \ref{eq4}, Eq.~\ref{eq13} is simplified as

\begin{equation}\label{eq14}
\begin{split}
F_{int} = L_x L_z \gamma (\theta) / \sin \theta + L_z (L_y +L_x/\tan \theta) \gamma ^{(+)} (|H_1|)+ \\
L_z(L_y-L_x/\tan \theta)\gamma^{(+)}(-|H_1|)+
2L_z \tau(|H_1|,\theta)
\end{split}
\end{equation}

We remark that the chosen geometry does not fix the angle $\theta$ a priori (unlike the case of the study by Mon et al. \cite{51,52} where free surfaces and boundary fields were avoided, and rather a screw periodic boundary condition in x-direction was used, that creates a fixed tilt angle, which together with the APBC in y-direction then ensures the presence of a tilted planar interface and full translational invariance). As a result, we can impose the condition that the contact angle $\theta$ is chosen such that $F_{int}$ takes a minimum, which yields (using also the abbreviation $\Delta f_1$ of Eq.~\ref{eq8}.)

\begin{eqnarray}\label{eq15}
(\frac {\partial F_{int}} {\partial \theta})_{H_1} = 0 &=& \frac {L_xL_z}{\sin ^2 \theta} \{ \gamma ' (\theta) \sin \theta - \gamma (\theta) \cos \theta + \Delta f_1\}\nonumber \\
&+& 2L_z(\partial \tau/\partial \theta)_{H_1}
\end{eqnarray}

In the thermodynamic limit $L_x \rightarrow \infty$, $L_z \rightarrow \infty$, the correction due to the line tension can be neglected, and we arrive at the well-known \cite{14,15,16,17} modified Young equation $\{\gamma ' (\theta) \equiv d \gamma (\theta) /d \theta)\}$

\begin{equation}\label{eq16}
\gamma (\theta) \cos \theta - \gamma ' (\theta) \sin \theta = \Delta f_1
\end{equation}

For a chosen value of $|H_1|$, the angle $\theta$ hence is an observable of the simulation, and although snapshot pictures reveal considerable fluctuations, one can sample well-defined magnetization profiles $m_k(y)$ in each (y,z) lattice plane parallel to the walls. Those layer magnetizations are constant away from the interface and show an inflection point when one crosses the interface. However, the details of these profiles do not matter, we simply need to know the constant values $m_{kx}^{(+)},m_{kx}^{(-)}$ away from the interface, and the average value

\begin{equation}\label{eq17}
\bar{m}_{kx} = \frac {1}{L_y}\sum \limits _{y=1} ^{L_y} m_{kx}(y)
\end{equation}

in the different layers. Then we can define distances $y_{kx}^{(+)}, y _{kx}^{(-)}$ of the dividing surface from the system boundaries as follows ($y_{kx}^{(+)} + y _{kx}^{(-)} =L_y$, per definition)

\begin{equation}\label{eq18}
\bar{m}_{kx} = \frac {1}{L_y}(y_{kx}^{(+)} m _{kx}^{(+)} + y_{kx}^{(-)} m _{kx}^{(-)})\;
\end{equation}

which readily yields the effective linear dimensions $y_{kx}^{(+)},\; y _{kx}^{(-)}$ of the coexisting domains in the respective (yz)-planes labeled by the index $kx$. From this analysis one can check that the interface indeed is planar (see Fig.~\ref{fig3}) , as hypothesized in Fig.~1, and from this purely geometrical method estimates of the contact angle $\theta$ as function of the field $H_1$ (and of temperature) can be extracted. Thus, this analysis defines the interface as a dividing plane between bulk phases, and defines the contact lines from the intersection of this plane with the surface planes. No assumption on the local structure of the interface near the surfaces needs to be made. Given the knowledge of the contact angle $\theta = \theta(H_1)$, at constant temperature we can transform Eq.~\ref{eq16} into a differential equation with respect to $H_1$ using $d\gamma/d \theta = (d\gamma/dH_1)(dH_1/d\theta)$, to obtain a differential equation for $\gamma$ as a function of $H_1$, which can be integrated numerically, using as an initial condition the fact that for $H_1=0$ we have a planar interface oriented perpendicular to the y-direction. As discussed amply in the literature (e.g. \cite{47,64}), numerous methods exist to estimate the interface tension of planar interfaces that are oriented perpendicular to one of the lattice axes of the simple cubic lattice. One of these methods is ``thermodynamic integration'' \cite{44,45}; we have already seen an application of this concept in Eq.~\ref{eq9}. for the sake of completeness, we give some comments on this method (as used in our work) in the following.

\subsection{Thermodynamic Integration}
As is well known \cite{44,45} the standard relation $\partial (\beta F)/\partial \beta =U$ between free energy $F$ and internal energy $U$ allows to estimate free energy differences via integration of $U$ with respect to inverse temperatures $\beta$,

\begin{equation}\label{eq19}
\beta F(\beta) - \beta _0 F(\beta_0) = \int \limits _{\beta _0}^\beta U(\beta ') d\beta '.
\end{equation}

While $U$ is a straightforward observable in Monte Carlo simulations, we need to specify a reference state at $\beta _0$ where $F(\beta_0)$ is known. In the low temperature phase, one wishes to use the information on the ground state properties, since $F(\beta\rightarrow \infty) = U(\beta \rightarrow \infty)$. Since at low temperatures $U(\beta)$ differs from the groundstate energy $U_0=U(\beta \rightarrow \infty)$ by terms proportional to $\exp (-6 \beta J)$ in the bulk and $\exp (-4\beta J)$ at a planar interface, it suffices to choose $\beta _0$ large enough such that these terms are negligible, e.g. $\beta_0J=10$.

In practice, what we actually need to do is not the computation of the free energy of a three-dimensional bulk system, but we rather consider the free energy difference $\Delta F$ of two systems with identical linear dimensions, the one system with APBC in y-direction, so that there occurs an interface oriented in the xz-plane, and the other system has PBC in y-direction. In z-direction we have PBC in both cases, and in x-direction we use free surfaces (without boundary fields, so $\theta = \pi/2$ in Fig.~1, and no GAPBC rather than APBC in y-direction are needed). The resulting free energy difference then yields directly the interface tension $\gamma (\pi/2)$, plus the corresponding line tension correction

\begin{equation}\label{eq20}
\Delta F = F_{APBC} - F_{PBC} = L_z L_x \gamma (\frac \pi 2) + 2L_z \tau(0,\frac \pi 2)
\end{equation}

Since for $\beta < \beta_c (\beta _c = 1/k_BT_c)$ both $\gamma(\pi/2)=0$ and $\tau(0,\pi/2) =0$, because there occur no interfaces for $T>T_c$, it also is convenient to use a reference state at $\beta_0=0.2$, for instance.

Using then $\Delta F/L_zL_x = \gamma(\pi/2) + (2/L_x)\tau(0,\pi/2)$ for different choices of $L_x$, it is straightforwardly possible to estimate both $\gamma(\pi/2)$ and $\tau(0,\pi/2)$.

We now turn to the interfacial excess free energies in the presence of the surface field $H_1$. Returning to Eq.~\ref{eq14}, we consider now the difference in the total interfacial energy as found in Eq.~\ref{eq14}, and the corresponding result for $\theta = \pi/2$, which is the free energy difference $\Delta F$ as found in Eq.~\ref{eq20}, plus a term $2L_xL_y \gamma^{(+)}(0)$ from the free boundaries (remember that $\gamma^{(+)} (0) = \gamma ^{(-)}(0)$, of course). This difference then is

\begin{equation}\label{eq21}
\begin{split}
F_{int} (H_1) - F_{int} (0) = L_x L_z [\gamma (\theta) / \sin \theta -\gamma (\pi/2)] + \\ L_z L_y [\gamma ^{(+)} (|H_1|) +
\gamma^{(+)}(-|H_1|)-2 \gamma^{(+)}(0)] + \\L_z L_x[\gamma ^{(+)} (|H_1|) - \gamma^{(+)}(-|H_1|)] / \tan \theta + \\2L_z [\tau(|H_1|,\theta)-\tau(0,\pi/2)]
\end{split}
\end{equation}

We note that a significant part of this expression, namely all interfacial excess terms associated with the free surface where the surface fields act, are readily found making use of Eq.~\ref{eq7} once more,

\begin{equation}\label{eq22}
\begin{split}
F_{int}(H_1) - F_{int} (0) - L_xL_z [\gamma(\theta)/\sin \theta - \gamma(\pi/2)]=\\ -\int \limits _0^{H_1} [m_1(H'_1 ) +m_n(H'_1)] dH_1'
\end{split}
\end{equation}

In this context, it does not matter that in the geometry of Fig.~1 the local layer magnetization $m_1(-|H_1|,y)$ and $m_n(|H_1|,y)$ are inhomogeneous as function of $y$, Eq.~\ref{eq7} remains valid also for a case of two-phase equilibrium in the system. As described above, the angle $\theta$ is found from direct observation, $\Delta f_1(|H_1|) $ is found from thermodynamic integration for boundary conditions where two-phase coexistence is avoided, and $\gamma (\theta)$ is found from exploiting the differential equation Eq.~\ref{eq16}, as described above. Hence it follows that the only unknown term on the left hand side of Eq.~\ref{eq22} then is $2L_z[\tau (|H_1|,\theta)-\tau (0,\pi/2)]$, which hence can be estimated when we compute the right hand side of Eq.~\ref{eq22} by thermodynamic integration.

Of course, the line tension term is down by a factor $1/L_x$ in comparison with the contribution of the (tilted) interface (cf.~Eq.~\ref{eq20}). As a consequence, one must exert great care in ensuring excellent accuracy when carrying out the various thermodynamic integrations. In addition, it is mandatory to repeat the computations for several choices of the linear dimensions $L_x,L_y$ and $L_z$ (which all have to be very large, in comparison with the lattice spacing), to make sure that the analysis is not hampered by finite size effects. We also draw attention to the fact that varying $L_z$ together with both $L_x$ and $L_y$ such that $L_xL_z = \textrm{const} $ and $L_yL_z = \textrm{const}$ the only variation predicted by Eqs.~\ref{eq20} - \ref{eq22} are proportional to $L_z$ and involve line tension contributions. E.g., a set of possible reasonable linear dimensions with $L_xL_z=4608$ and $L_yL_z=27648$ is $L_x =96, \; L_y=576, \; L_z=48; \; L_x=48, \; L_y=288, \; L_z = 96$; $L_x=24,\; L_y=144,\; L_z=192$. Carrying out the thermodynamic integrations for such sets of linear dimensions one can test the consistency of the estimates for the line tension; note that for this quantity no other calculations in the literature are available as yet.

\section{Numerical Results}
\subsection{Angle-dependent interfacial tension}

As described in Sec.~IIB, we proceed by varying $H_1$ from $H_1=0$ up to a value that is still somewhat smaller than the field $H_{1c}(T)$ where critical wetting would occur, using the geometry of Fig.~1, in order to estimate the quantities $m_{kx}^{(+)},m_{kx}^{(-)}$ and $\bar{m}_{kx}$ \{Eqs.~\ref{eq17},\ref{eq18}\} from which the estimate of the contact angle $\theta$ is extracted (Fig.~4a). Of course, symmetry requires that for $H_1=0$ the contact angle $\theta = \pi/2$, irrespective of temperature; for the chosen linear dimensions, we can follow $\theta$ down to about $\theta \approx 20 ^\circ$, before strong fluctuations due to the proximity of the critical wetting transition (which the chosen model exhibits \cite{55,58}) would make the estimation unreliable.

While this estimation uses the GAPBC where an interface runs across the thin film, as indicated in Figs.~1,3, we also carried out simulations with simple PBC, where the thin film (for $H_1 < H_{1c}(T))$ is in a state which is uniform in y-direction, exhibiting a magnetization profile across the film, as shown schematically in Fig.~2. Using in this geometry the thermodynamic integration, Eq.~\ref{eq9}, we can estimate the contact angle $\theta$ also from Eq.~\ref{eq8}, using for $f_{int}(T) $ the interfacial energy for planar interfaces in the Ising model, as estimated by Hasenbusch and Pinn \cite{50}. Using also thermodynamic integration with respect to inverse temperature \{Eqs.~\ref{eq19},\ref{eq20}\}, we have obtained estimates for $f_{int}(T)$ as well, and found excellent agreement with the previous work \cite{50}. Thus the use of these estimates for $f_{int}(T)$ in Eq.~\ref{eq8} causes only negligible errors; hence the discrepancies between contact angles $\theta$ (Fig.~4b) estimated from the standard Young's equation, Eq.~\ref{eq8} and our more general direct method clearly are statistically significant, and must be attributed to the anisotropy of the interfacial tension $\gamma(\theta)$, giving rise to a modified Young equation, Eq.~\ref{eq16}. The comparison shown in Fig.~4b shows that the discrepancies are largest at low temperatures, but near the critical temperature (e.g. for a reduced temperature $T=3.8$) these discrepancies are already essentially negligible. However at $T=3.0$ there are still systematic discrepancies visible; this fact causes some doubt in the accuracy of the analysis of wall-attached droplets due to Winter et al. \cite{33,34}, who did rely on the accuracy of Eq.~\ref{eq8} at this temperature.

In Sec.~II B it was also pointed out that the function $\theta = \theta (H_1)$ can be used together with Eq.~\ref{eq16} to integrate this differential equation and obtain $\gamma(\theta)$ explicitly. The results are displayed in Fig.~5. As it must be, the anisotropy is maximal for $\theta = \pi/4$. Fig.~6 compares the results for this maximum anisotropy with previous estimations \cite{52,65}. For $T >T_R$ the anisotropy is rather small, it is only of the order of few percent; thus an extremely good statistical accuracy of the original simulation data is stringently required for meaningful results. Thus, the excellent agreement of our estimation with the results from Bittner et al. \cite{65} is very gratifying.

At low temperatures the anisotropy is much larger; note that at $T=0$ an interface inclined under an angle $\theta =\pi/4$ involves two broken bonds per spin in the interface, but the interfacial area also is enhanced by a factor $\sqrt{2}$, and hence the value of $\gamma (\pi/4)/\gamma(\pi/2)=2/\sqrt{2}=\sqrt{2}$ at $T=0$. Previous estimates, due to Mon et al. \cite{52} 25 years ago, suffer from significant statistical errors (particularly near $T=T_R$) and also from finite size effects, since the chosen linear dimensions $L=32$ and $L=48$ clearly are insufficient for $T$ near $T_R$, where these data hence suffer from finite size rounding. While the data by Mon et al. \cite{52} were obtained from thermodynamic integration with $T=0$ as a reference point, and using staggered APBC where $\theta$ was fixed by the geometry, Bittner et al. \cite{65} used estimations of the minimum in the probability distribution $P_L(m=0)$ of the magnetization $m$ relative to the maximum $P_L(m=\pm m_{coex})$, using staggered periodic boundary conditions which enforce for $m=0$ a domain configuration with domain walls inclined under the angle $\theta$. Thus, all the estimations compared here have used different methods.

\subsection{Interfacial stiffness}
For $T > T_R$ theory \cite{31,49,66} predicts that $\gamma(\theta)$ at $\theta = \pi/2$ has a quadratic expansion

\begin{equation}\label{eq23}
\gamma (\theta) = \gamma (\frac \pi 2) [1+c (\frac \pi 2 - \theta)^2 + \ldots ] \quad ,
\end{equation}

unlike the regime $T <T_R$ a linear term in $|\frac \pi 2- \theta|$ is present. Since the interfacial stiffness is defined as \cite{66}

\begin{equation}\label{eq24}
\kappa = \gamma (\pi/2) + \gamma ^{''} (\theta = \pi/2) = \gamma (\pi/2) [1+2c],
\end{equation}

we can use our results for $\gamma (\theta)$ to extract estimates for $\kappa$ (at different temperatures) from our data. Numerically this is a somewhat delicate task, as discussed in more detail in \cite{67}; but again our estimation is in reasonable agreement with previous work \cite{50}, Fig.~7, where $\kappa$ was extracted from an analysis of capillary wave induced interfacial broadening. As is well-known, (see e.g. \cite{37,48,49,54,66}), $\kappa$ (rather than $\gamma(\frac \pi 2)$) enters as a ``coupling constant'' in the capillary wave Hamiltonian that describes the free energy cost of long-wavelengths interfacial fluctuations; therefore a precise estimation of the temperature dependence of $\kappa$ is of great interest (Fig.~7). It is important to recall that $\kappa$ is singular when the temperature $T$ approaches the roughening transition $T_R$ \cite{30,31,68}

\begin{equation}\label{eq25}
\kappa (T) \beta = \frac \pi 2 [1-C(\frac {T-T_R}{T_c})^{1/2} + \ldots]\quad ,
\end{equation}

where the constant $C$ has been estimated as \cite{68} $C=1.57 ß\pm 0.07$. Fig.~7b shows that our data occur ``in the right ball park'', but clearly the singularity at the roughening transition is rounded off. In fact, since the roughening transition is accompanied by an exponentially strong divergence of the correlation length describing interfacial height fluctuations and their correlation, strong finite size rounding is expected, despite the large linear dimensions used. Of course, the constant $C$ (which relies on an estimate of the step free energy for $T \rightarrow T_R$ \cite{51,52}) is not really known accurately either; thus another estimate for $C$ (resulting from a fit of Eq.~\ref{eq25} to our three data points at $T=2.55, 2.60, 2.65)$ is shown for comparison. However, at the shown temperatures neglected higher order terms in the expansion quoted in Eq.~\ref{eq25} also might matter. Thus, a high precision study of the singular behavior of $\kappa$ near $T_R$ must be left to future work.

\subsection{Step Free energy}
When one studies tilted interfaces at temperatures $T$ well below the interfacial roughening transition temperature $T_R$, the microscopic picture of an interface configuration is a ``staircase'' of terraces with steps that have somewhat irregular edges (Fig.~8). At a standard step, the height of the terrace (relative to the x-z plane in Fig.~1, i.e. the y-coordinate of the terrace plane) increases by one lattice unit. Of course, these steps are one-dimensional objects, and run strictly straight along the z-direction in Fig.~1 at $T=0$ only; at any nonzero temperatures these steps exhibit kinks (forward and backward in y-direction), giving rise to a rather irregularly fluctuating local width of the terraces as well. In addition to this description of an interface in terms of the well-known terrace-step-kink model of surface physics \cite{69,70,71}, there also occur localized clusters of terrace height fluctuations (up or down) involving only a few neighboring lattice sites (or even only single sites, visible as small cubes in Fig.~8b); such localized defects need more energy, and hence are an insignificant detail, that can be ignored in the following. As is well known, for $T <T_R$ interfaces that coincide with the simple xz lattice plane in Fig.~1 would be perfectly flat, if such localized clusters of height fluctuations are ignored. Tilting such an interface by a very small angle then costs a free energy $L_z L_x f_s(T)(\frac \pi  2- \theta)$ in Fig.~1, where $f_s(T)$ can be interpreted as the free energy cost (per lattice site) to create a single step on an otherwise planar interface, i.e.

\begin{equation}\label{eq26}
f_s(T) = |\frac {d \gamma (\theta)}{d \theta}|_{\theta = \pi/2} \quad .
\end{equation}

Above the roughening transition temperature, $f_s(T>T_R)\equiv 0$, as is also implied by Eq.~\ref{eq23}, and thus the roughening transition temperature can also be characterized by the vanishing of $f_s(T)$ \cite{30,31}. Actually it is predicted that near $T_R \quad f_s(T)$ scales proportional to the inverse of the correlation length $\xi_R$ of interfacial height fluctuations near the roughening transition \cite{31}

\begin{equation}\label{eq27}
\xi_R(T) \propto \exp \{A[1-T/T_R]^{-1/2}\}\quad ,
\end{equation}

where $A$ is a constant. However, at the outset we stress that the relation $f_s(T) \propto 1/\xi_R(T)$ is extremely difficult to verify in simulations, since then the linear dimensions $L_x,L_z$ must be chosen much larger than $\xi_R (T)$ itself, of course.

We again proceed as in Sec.III.A where the contact angle $\theta$ is obtained from the geometric construction in terms of $\bar{m}_{kx},m_{kx}^{(+)}$ and $m_{kx}^{(-)}$ \{Eqs.~\ref{eq17}, \ref{eq18}\}, varying the strength of the surface field $H_1$, but now we focus on the behavior at low temperatures (Fig.~9). One sees that there exists a range of surface fields up to some critical field $H_1^*$ for which the contact angle stays at 90$^\circ$ before it starts to decrease; this critical field $H_1^*$ increases with decreasing temperature. Of course, only when this critical field is exceeded are surface steps formed.

In order to understand the existence of this critical field $H_1^*$ in this kind of simulation set up qualitatively, we formulate a simple model of a tilted interface at very low temperatures. We assume that on the length $L_x$ there either is no step or a single step (which then involves a free energy cost $f_s(T)L_z)$. This step can go in the positive (+1) or in the negative (-1) y-direction, and the unnormalized weights of the three possibilities are (we assume that the local magnetization at the walls is $\pm 1$ at low temperatures)

\begin{equation}\label{eq28}
W_0= 1 \quad \textrm{(no step)}
\end{equation}

\begin{equation}\label{eq29}
W_{+1} = L_x \exp (- \frac {f_s(T)L_z}{k_BT})\exp (\frac {2H_1L_z}{k_BT})
\end{equation}

\begin{equation}\label{eq30}
W_{-1}=L_x \exp (- \frac {f_s(T)L_z}{k_BT}) \exp(-\frac {2H_1L_z}{k_BT})
\end{equation}

The average inclination angle of the interface then is

\begin{equation}\label{eq31}
\langle \phi \rangle = \frac \pi 2 - \theta = \frac {1}{L_x} (W_{+1} - W_{-1})/Z, \quad Z=W_0 + W_{+1} + W_{-1}
\end{equation}

From these equations, one sees that $\langle \phi \rangle$ will be essentially zero up to

\begin{equation}\label{eq32}
H_1^* \approx \frac 1 2 f_s (T) - \frac {1} {2L_z} k_BT \ln L_x
\end{equation}

Of course, this consideration can be generalized to include more steps, but the conclusion that for $L_z \rightarrow \infty$ the onset of a nonzero angle $\langle \phi \rangle$ occurs for $H_1^* \approx f_s(T)/2$ is not altered. Fig.~10 shows that Eq.~\ref{eq31} predicts a far too rapid decrease of $\theta$ when $H_1^*$ has been passed, at most temperatures of interest; as is obvious from the approximations made, the treatment disregards all the effects of kinks at the steps (which are expected to matter, see the snapshot pictures, Fig.~8). The finite size effects are relatively large; however, one should remember that the kinks along a step at a surface are to be compared to the kinks (domain walls) in the one-dimensional Ising model, leading to a correlation length $\xi_{1 d} \approx -1/\ln \{\tanh (J/k_BT)\} \approx \frac 1 2 \exp(2J/k_BT)$ at low temperatures. These kinks of the steps lead to lateral displacements of the steps at the terraces, which add up in a random walk like fashion, leading to a mean-square width $w^2_y= a^2L_z/\xi_{1d}$ over which the steps ``wander'' on the terrace in y-direction (cf. also Fig.~8b for an illustration). This effect clearly leads to effective interactions between steps \cite{48}, which are out of consideration here.

When we disregard the problem that variation of $\theta$ with $H_1$ beyond $H_1^*$ cannot be described, but simply rely on the estimate $f_s(T)= 2H_1^*$, we can fit the constant $A=(2C/\pi) \sqrt{T_c/T_R} $ in Eq.~\ref{eq25} which follows from the result that $f_s(T) \propto 1/\xi_R(T)$. The results shown in Fig.~11 are in qualitative agreement with the earlier estimates of Mon et al. \cite{51,52} that were obtained by a different method as mentioned above. However, the latter authors attempted a tentative finite size extrapolation that yielded significantly smaller values for $f_s(T)$ near $T_R$ than the present data since their data exhibit pronounced finite size effects (see Fig.~\ref{fig11}). As we here have not attempted to carry out a finite size scaling analysis of the critical behavior associated with the roughening transition, in view of the massive effort in computational resources that would be required for an adequate study along such lines, this remains an open problem.

\section{A Study of the line tension}
Estimation of the line tension of the Ising model (with unchanged exchange interactions on the free surfaces, but varying the strength of the surface field) was one of the main objectives of the present investigation. We start out with the case $H_1=0$, using Eqs.~\ref{eq19}, \ref{eq20}. Fig.~12 shows the result for $\tau (0,\pi/2)$ as a function of temperature, from the ground state up to the critical point. As one can verify simply by bond-counting, for an interface hitting a free surface at $T=0$ the contact line does not create any extra energy cost (or gain, respectively) in the ground state. The line tension normalized by temperature) is found to be negative throughout, and reaches a minimum of about $-0.25 $ at about T = 3.0, and is close to - 0.20 J at about T = 4.0. These findings are compatible with previous estimates by Winter et al. \cite{33,34} at those temperatures. As expected, $\tau(0,\pi/2)$ vanishes again as the bulk critical temperature is approached. A hyperscaling argument implies that $\tau(0,\pi/2)\propto \xi_b^{-1}$ where the bulk correlation length varies as $\xi_b \propto (1-T/T_c)^{-\nu_b}$ with \cite{72} $\nu_b\approx 0.63$. While our data are roughly compatible with this estimation, significantly larger systems (and more statistical effort) would be required to verify this hypothesis with significant precision.

We now turn to the estimation of the line tension at nonzero surface fields, using Eqs.~\ref{eq21}, \ref{eq22} and making use of the possibility of choosing linear dimensions $L_x,L_y$ and $L_z$ such that the areas $L_xL_z$ and $L_yL_z$are kept constant when $L_z$ is varied, in order that the coefficient of $L_z$ (which is related to the line tension contribution, as discussed in Sec.II.C) can be isolated. Fig.~13 presents some examples for the application of this method. It is clearly seen that linear dimensions $L_z=16, L_z = 32$ are systematically below the asymptotic straight line, but the three larger systems (with $L_z$ = 48, 64 and 96) seem to yield reliable estimates. Note that this calculation is not expected to give reliable results either near $T_R$ or near $T_c$, since the film thickness $L_x$ for $L_z = 96$ (when also $L_y = 96$, while for the other choices $L_y$ is always larger) is only $L_x=16$, and hence finite size effects must be expected. On the other hand, if we would have chosen all linear dimensions larger ,the signal to noise ratio for the line tension effect would have been even more unfavorable.

Fig.~14 shows then our final results for the variation of the line tension as a function of contact angle for various temperatures. Normalizing the line tension by the exchange constant $J$ (rather than temperature) the scale of $\tau(H_1,\theta)$ is of order unity, as expected, and the most negative values now are found for $T=3.5$ (rather than by $T=3.0$, which is found when one normalizes by $T$, cf. Fig.~12). We expect that the line tension at all temperatures should vanish when the critical wetting transitions $(\theta \rightarrow 0$) is approached, but in the range $0<\theta<30^\circ$ no reliable data as yet have been generated.


The present work is the very first attempt to estimate the line tension in the Ising model over a wide range of temperatures and contact angles, and thus there are no data to compare our results with, apart from those of Winter et al. \cite{33,34} extracted from the analysis of wall-attached sphere-cap shaped small droplets. These authors predicted a linear increase of $\tau (H_1,\theta)$ with $|H_1|$ at T = 3.0. Since the contact angle $\theta$ varies linearly with $|H_1|$ near $\theta = \pi/2$, these data imply a linear increase of $\tau$ with $\theta$ as well, which is not observed here. However, the analysis of Winter et al. \cite{33,34} was based on several assumptions, in particular the anisotropy of the interfacial tension was neglected, which is not warranted at $T=3.0$ in the light of results in Sec.~III.A, and hence the free energy barrier of the droplet in the bulk is already higher than expected for a spherical droplet, consistent with a recent study \cite{35}.

\section{Conclusions}
This work has presented a joint study of the anisotropic interfacial free energy and the dependence of the line tension of the Ising model on temperature and contact angle, using Monte Carlo simulations of thin Ising films with free surfaces at which surface fields of opposite sign but equal absolute magnitude act. It is shown that a generalized antiperiodic boundary condition (GAPBC), Sec.~IIA, Eqs.~\ref{eq10}, \ref{eq11} allows the simulation of a single interface running across the film, if the magnitude of the surface field is chosen to fall in the regime of partial wetting (for the corresponding semi-infinite Ising system), respecting the full translational invariance in the directions parallel to the free surfaces. Using these GAPBC, the total surface and interfacial excess free energy contributions are analyzed (Sec.~II.B, Eqs.~\ref{eq13}, \ref{eq14}), making use of the special spin reversal symmetries of the Ising model. Minimizing this free energy with respect to the contact angle $\theta$ leads to a new differential equation for the anisotropic interfacial tension $\gamma (\theta)$, involving a line-tension correction \{Eq.\ref{eq15}\}. In the thermodynamic limit, the result coincides with the well-known \cite{14,15,16,17} modified Young equation \{Eq.~\ref{eq16}\}. It then is shown how the contact angle can be inferred from observations of the local magnetization in the rows parallel to the walls \{Eqs.~\ref{eq17}, \ref{eq18}\}, and $\gamma(\theta)$ then is found explicitly from numerical integration, using the observed function $\theta = \theta (H_1)$. The line tension $\tau$ for the case of $\theta = \pi/2$ (when no surface fields act) is extracted from thermodynamic integration with respect to inverse temperature \{Eqs.~\ref{eq19},\ref{eq20}, Fig.~12\}, and studying the $L_z$-dependence of the corresponding excess free energy (for the geometry of Fig.~1) for a situation when the surface area $L_yL_z$ and the interface area $L_xL_z$ are kept constant, the dependence on the line tension $\tau$ on both contact angle and temperature is obtained (Fig.~14).

Our results on $\gamma(\theta = \pi/4)$ compare very well with previous work by Bittner et al. \cite{65}, who had used a different method (Fig.~6). From the results for $\gamma(\theta)$ near $\theta=\pi/2$, we extract the interfacial stiffness (Fig.~7) and find its temperature dependence in reasonable agreement with the results of Hasenbusch and Pinn \cite{50},  that were based on an analysis using capillary wave concepts. For low temperatures, the step free energy is obtained (Fig.~11) and compared to previous results of Mon et al. \cite{51,52}. Our data are qualitatively compatible with the singular behavior near the roughening transition temperature $T_R$ (Figs.~7,11), but a state of the art finite size scaling analysis due to critical slowing down would require a massive computational effort, and remains to be done.

 The computations analyzed here have required a large number of simulations at different values of the surface field, in order to reach the necessary statistical accuracy in the estimation of the contact angle and for the thermodynamic integration \{Eq.~\ref{eq22}\}. Since also rather large lattices need to be used, typically $184 \times 504 \times 504$ = 46 738 944 sites (Fig.~11), the availability of fast code implementations on GPUs (see Appendix) was crucial for the progress obtained in the present work.

 However, it is clear that also the present work still is an intermediate step only. In order to be able to carry out the Wulff construction \cite{18,19,20,21,22,23} to find the equilibrium shape of minority domains embedded in the majority phase, $\gamma(\theta)$ is not sufficient, and one needs to consider interfaces tilted with respect to the xz lattice plane by two angles, $\gamma(\theta,\varphi)$. Only the knowledge of this full function would allow to quantitatively account for the observed enhancement of the surface excess free energy of such minority domains over the simple estimation in terms of spherical droplets using the ``capillarity approximation'' \cite{4,5,6,7,8}, i.e. $4\pi (3V/4\pi)^{2/3} \gamma(0,0)$, V being the droplet volume. Schmitz et al. \cite{35} observed that the enhancement factor gradually grows from 1 for $T=T_c$ to $6/\pi$ at $T=0$. We have already pointed out, that a quantitatively reliable analysis of the behavior of $\gamma(\theta)$ near the roughening transition temperature also still remains to be done. A still more complicated step then is to address anisotropic surface tensions of crystal surfaces for off-lattice models. We hope that the present work will encourage all such extensions.

\underline{ Acknowledgements}: One of us (B. J. B.) is grateful to the Max Planck Graduate Center at Mainz for financial support and the Zentrum f\"ur Datenverarbeitung (ZDV) Mainz for computational resources.

\section{Appendix}

Our parallel GPU implementation of the Ising model is based on a checkerboard update, which is described in detail in Ref.~\cite{73} and based on previous implementations in Refs.~\cite{74,75}.
Most simulations were performed using the CUDA (compute unified device architecture) framework provided by Nvidia, but implementations using OpenCL have also been tested. 
%
For efficient sampling on the GPU it is advisable that computation is structured in thread-blocks with a constant number of threads. 
Each of the threads processes a three-dimensional array of 4$\times$4$\times$8=128 spins sequentially. Threads are typically organized in thread blocks of 8$\times$8$\times$8 threads for maximum performance,
but we have also undertaken simulation with blocks of 4$\times$4$\times$4 threads to allow for more flexibility in the choice of system sizes. Parallel execution of threads is scaled by CUDA to fit the GPU that is used.
%
For the periodic boundary conditions, an extra layer of spin blocks of size 4$\times$4$\times$8 is simulated at the two borders in each dimension with periodic boundary conditions.
This extra block is then copied to the other side of the lattice, before the next simulation step takes place, effectively implementing periodic boundary conditions.
For antiperiodic boundary conditions, the content of the block is modified before it is written back into the simulation lattice.
The combination of these restrictions leads to the unusual system sizes simulated. 6 thread blocks in x-dimension, e.g., therefore correspond to 6$\times$8 threads each containing 4 spins in x-direction (192 spins). From this number the
two border spin blocks, which are only used to mediate the boundary conditions, need to be subtracted. In this example the number of spins in x-direction is therefore reduced to 192-8=184 spins.

\newpage

\begin{figure}
\centering
\includegraphics [scale=0.28]{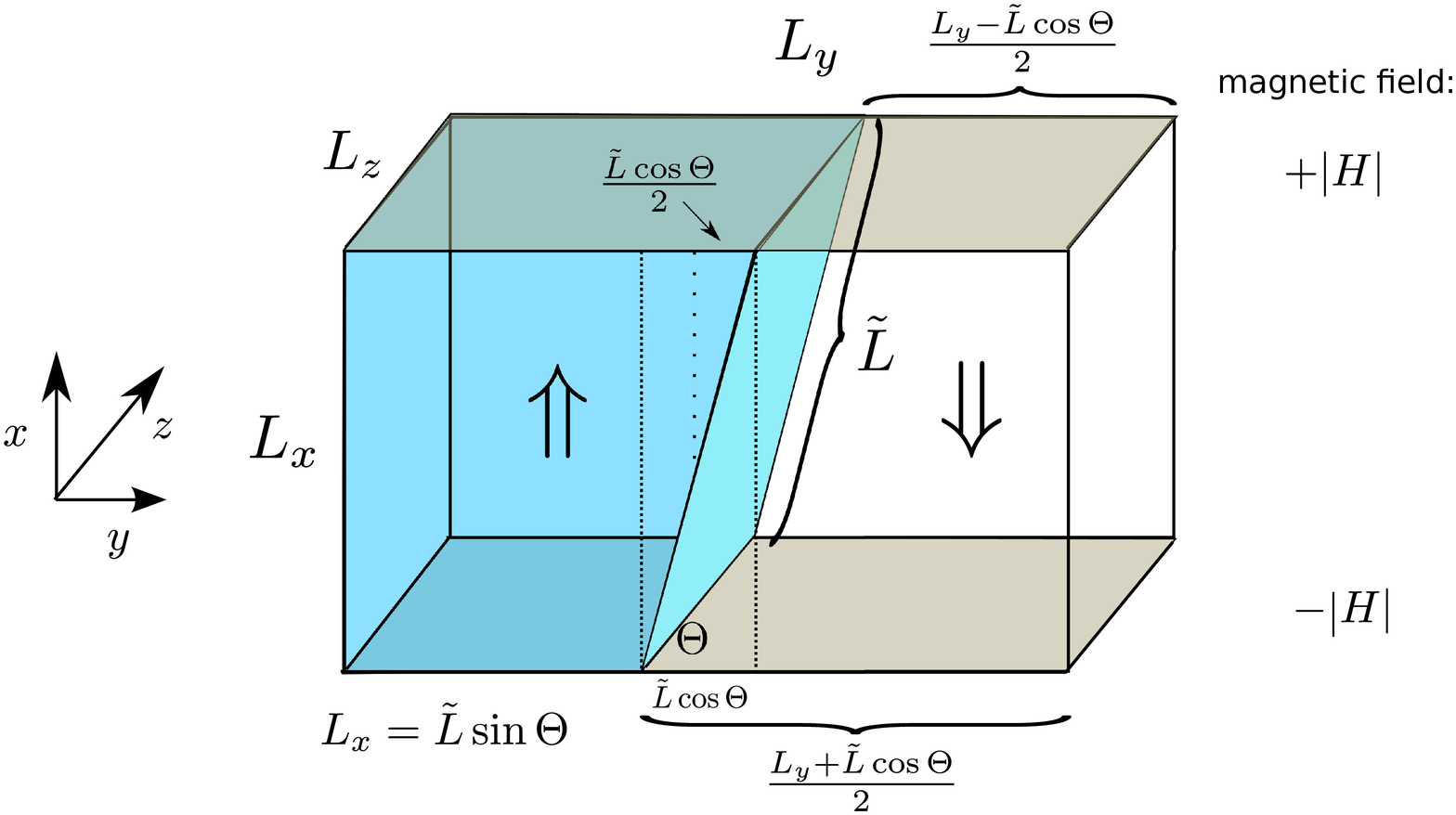}
\caption{\label{fig1} (Color online) Sketch of the simulation geometry, using a cubic lattice with simulation box  linear dimensions $L_x,L_y$ and $L_z$ in x,y and z directions. While a periodic boundary condition is used in z direction, in x-direction two free surfaces of linear dimensions $L_y \times L_z$ are used, at which boundary fields $H_1=-|H_1|$ at the bottom and $H_n=+|H_1|$ at the top act. When one uses a generalized antiperiodic boundary condition (GAPBC), as explained in the text, in the remaining y-direction, one can stabilize phase coexistence between two domains with opposite magnetization $+m_{coex}$, $-m_{coex}$ in the bulk (symbolized by the double arrows), separated by a single planar domain wall that is inclined by an angle $\theta$ with respect to the (yz) plane. For $H_1=0$ the Ising symmetry requires $\theta = \pi/2$ and the GAPBC reduces to the standard APBC.}
\end{figure}

\begin{figure}
\centering
\includegraphics [scale=0.60]{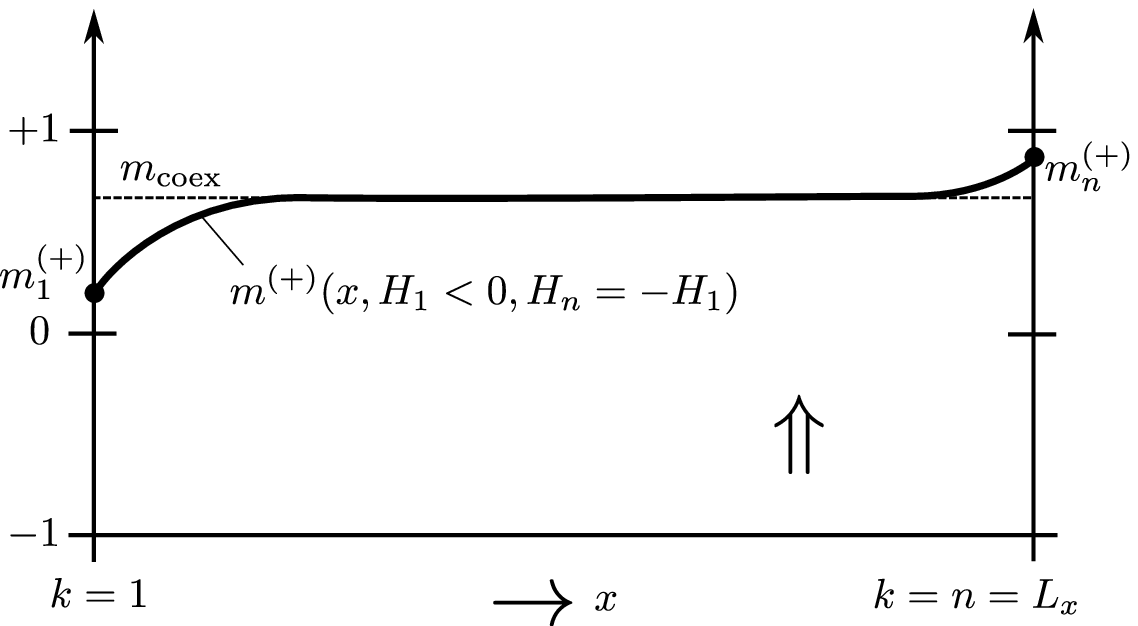}
\includegraphics [scale=0.60]{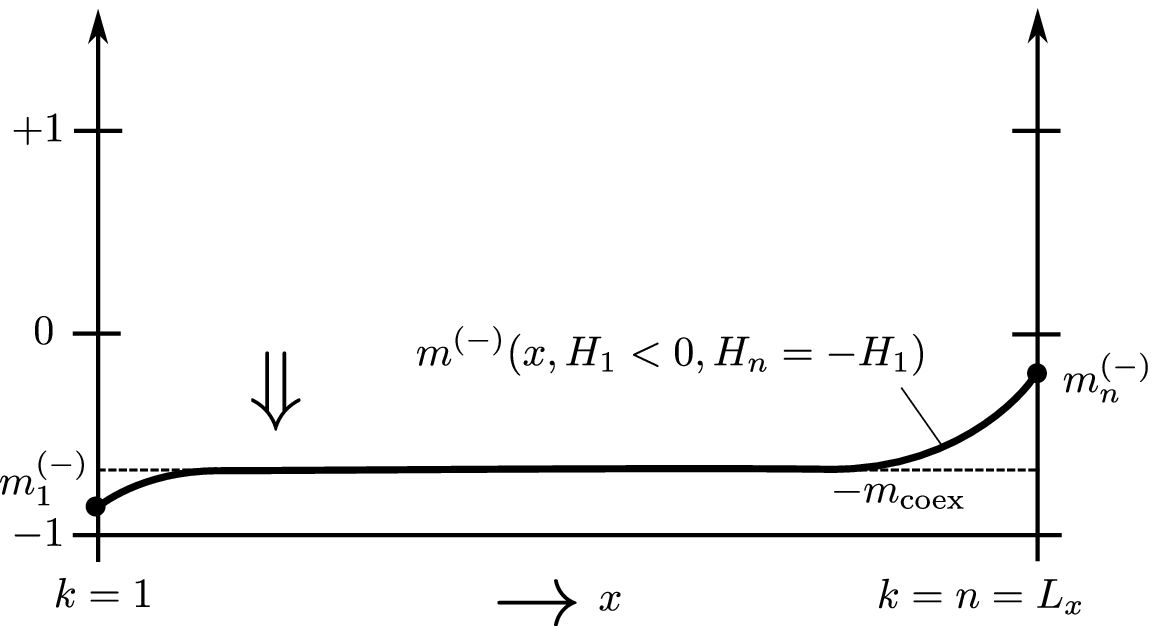}
\caption{\label{fig2} Schematic profiles of the layer magnetization $m^{(+)}(x)$, $m^{(-)}(x)$ across the film (with nonzero boundary fields $H_1<0, \; H_n=-H_1>0$) for the two coexisting phases (symbolized by the double arrows up or down, as in Fig.~1). For simplicity, x is treated as a continuous variable rather than the layer index $k=1,2,\ldots,n=L_x$. It is assumed (as in Fig.~1) that a state of partial wetting is realized. The symmetry relations Eqs.~\ref{eq5}, \ref{eq6}, that apply for these profiles, are a consequence of the Ising spin reversal symmetry. Note that the application of GAPBCs transforms the profile with positive total magnetization (upper part) to the profile with negative total magnetization (lower part) or vice versa.} 
\end{figure}

\begin{figure}
\centering
\includegraphics[scale=0.6]{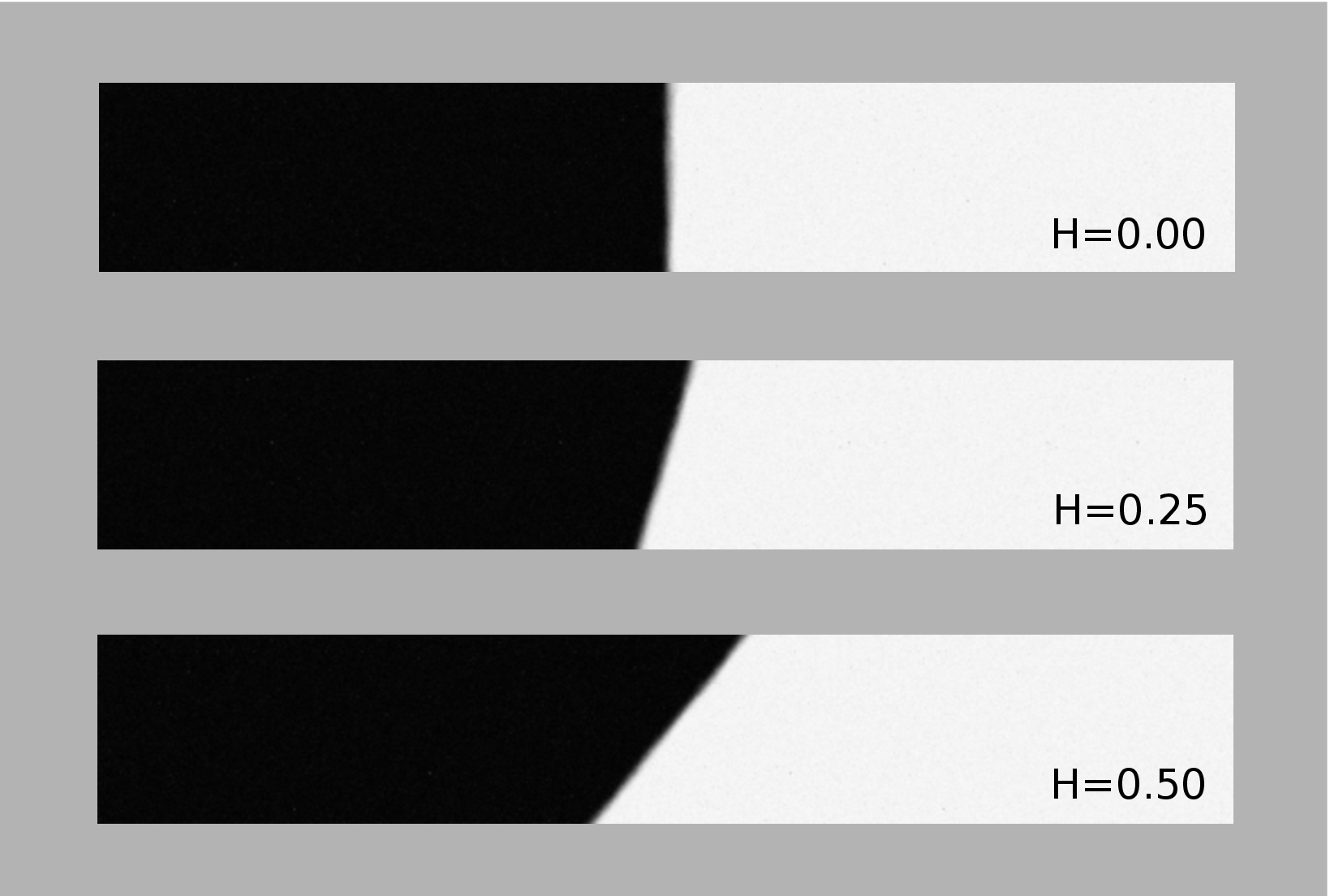}
\caption{\label{fig3} Configuration snapshots density profile integrated over the $z$-axis of a system as in Fig.~1, at $k_BT/J= 3.0$, with $L_x = 88, L_y= 504, L_z=504,$ for different surface fields $H_1$, projected into the xy-plane. Cases shown are $|H_1|/J= 0.00$ (a) ,$0.25$ (b), and $0.50$ (c). Up-spins are shown in dark, down-spins are not shown.}
\vspace{1cm}
\end{figure}

\begin{figure}
\centering
\includegraphics [scale=0.28] {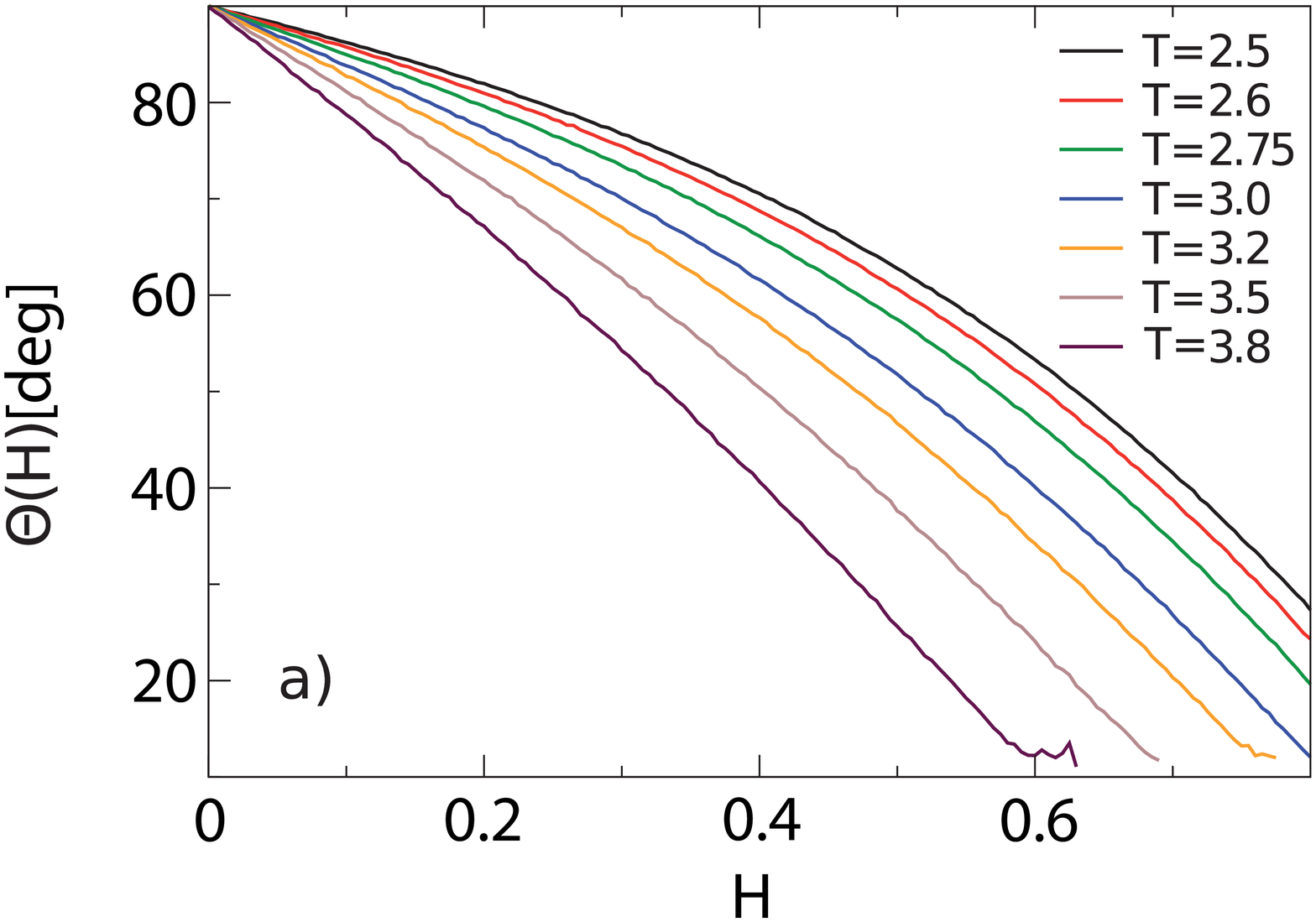}
\includegraphics [scale=0.28] {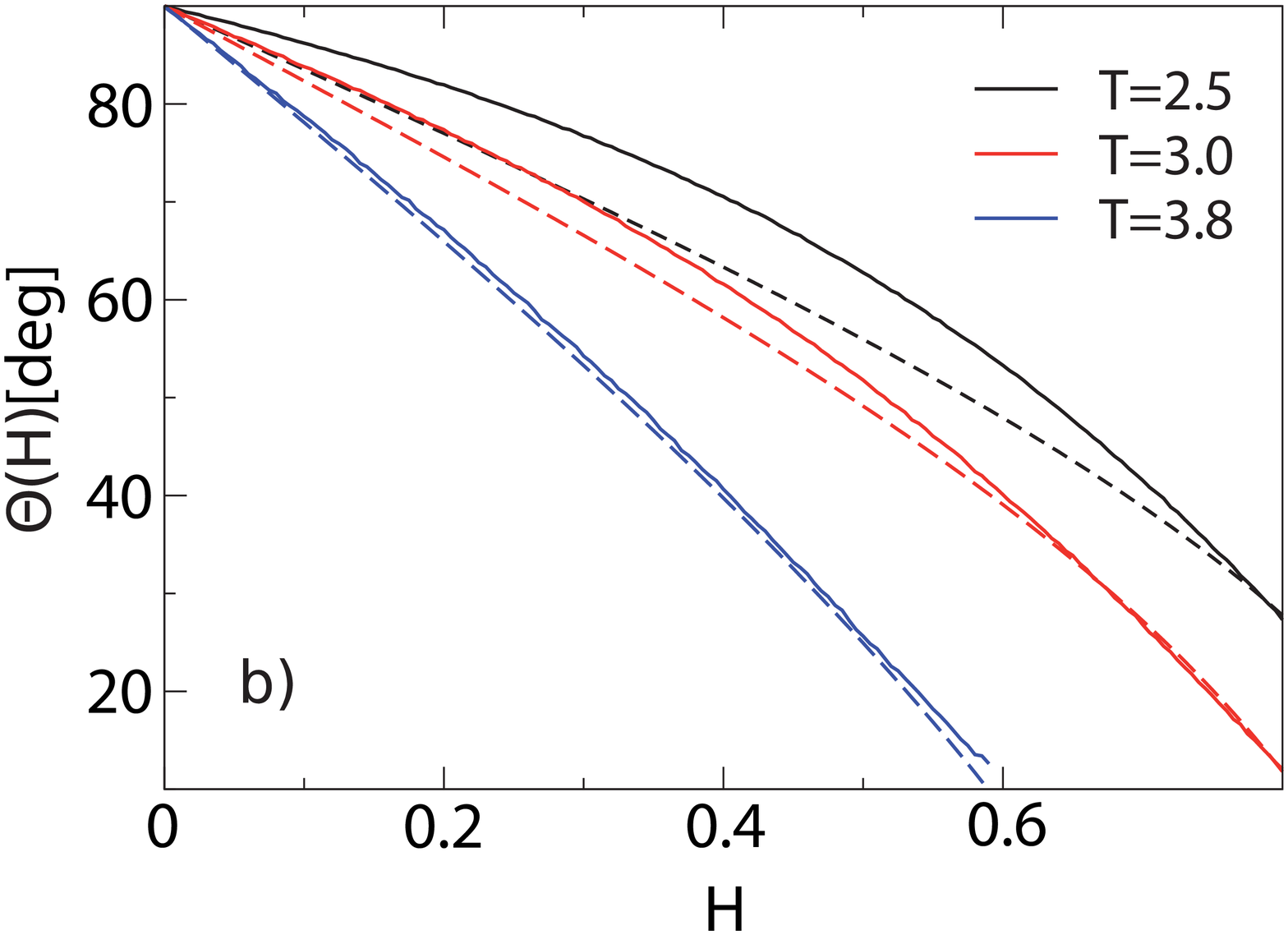}
 \caption{\label{fig4} (Color online) (a) Plot of the contact angle $\theta$ as a function of the surface field $|H_1|$ for different reduced temperatures $T$ (from top ($T=2.5$) to bottom ($T=3.8$), measured in units of $J/k_B$) as indicated.
(b) Same as (a), but comparing with the predictions based on the standard Young equation, Eq.~\ref{eq8} [broken curves]. Three reduced temperatures are shown: T = 2.5, 3.0 and 3.8 (from top to bottom).}
\end{figure}

\begin{figure}
\centering
\includegraphics [scale=0.37]{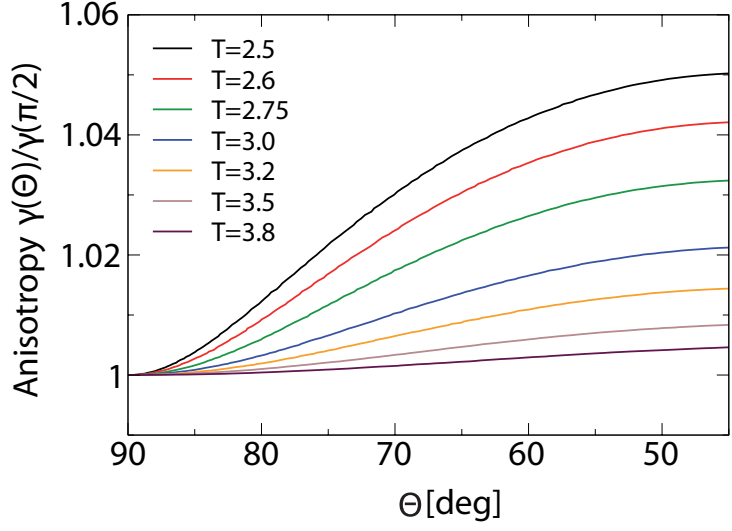}
\caption{\label{fig5} (Color online) Surface tension anisotropy $\gamma(\theta)/\gamma (\pi/2)$ plotted vs. $\theta$, for various reduced temperatures $T$ from $T=2.5$ (top) to $T=3.8$ (bottom), as indicated. Only the regime from $\theta = 90^\circ$ to $\theta = 45^\circ$ is shown, since on the simple cubic lattice a symmetry $\gamma (\theta) = \gamma (\pi/2-\theta)$ must hold.}
\end{figure}

\begin{figure}
\centering
\includegraphics [scale=0.37] {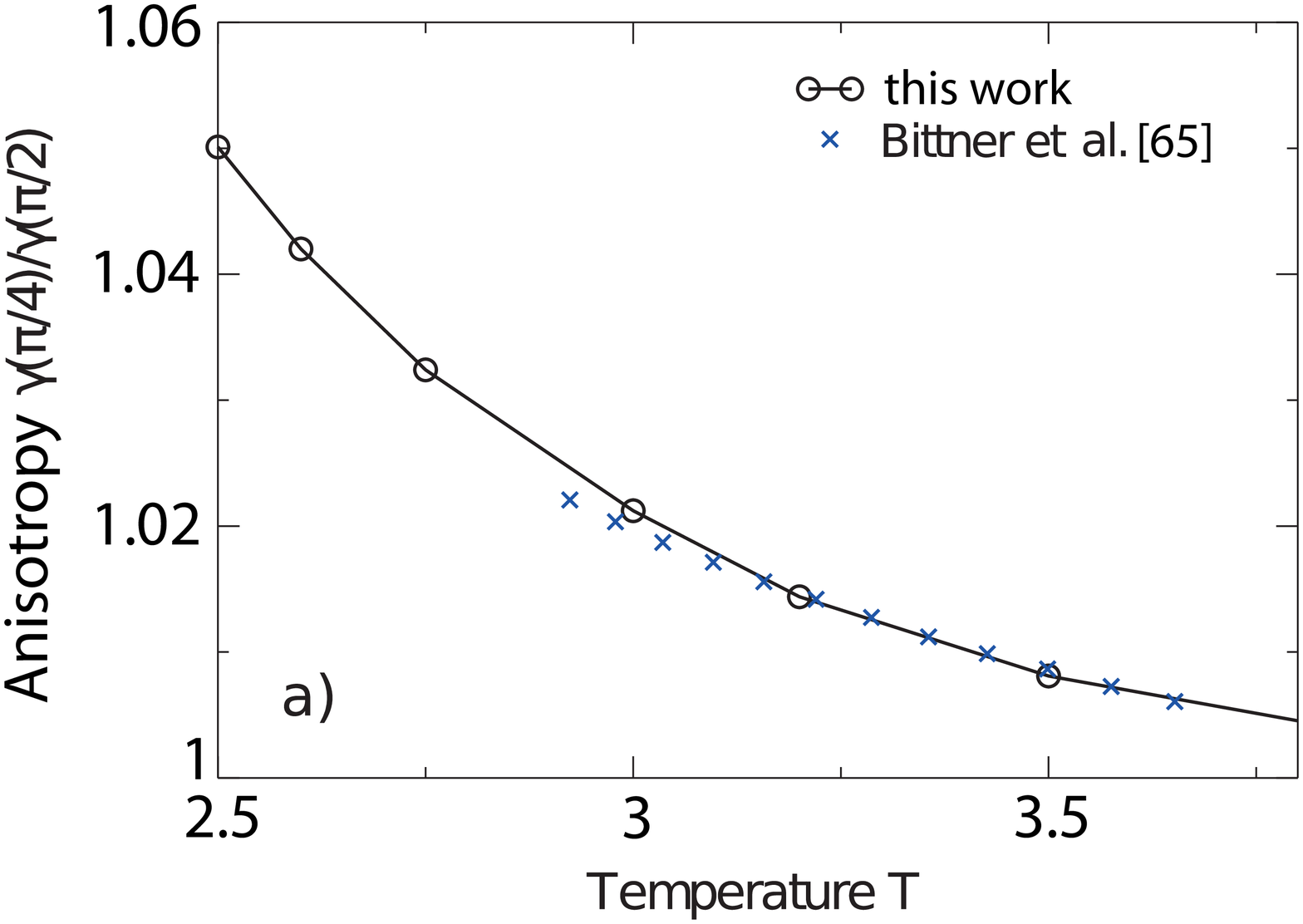}
\includegraphics [scale=0.37]{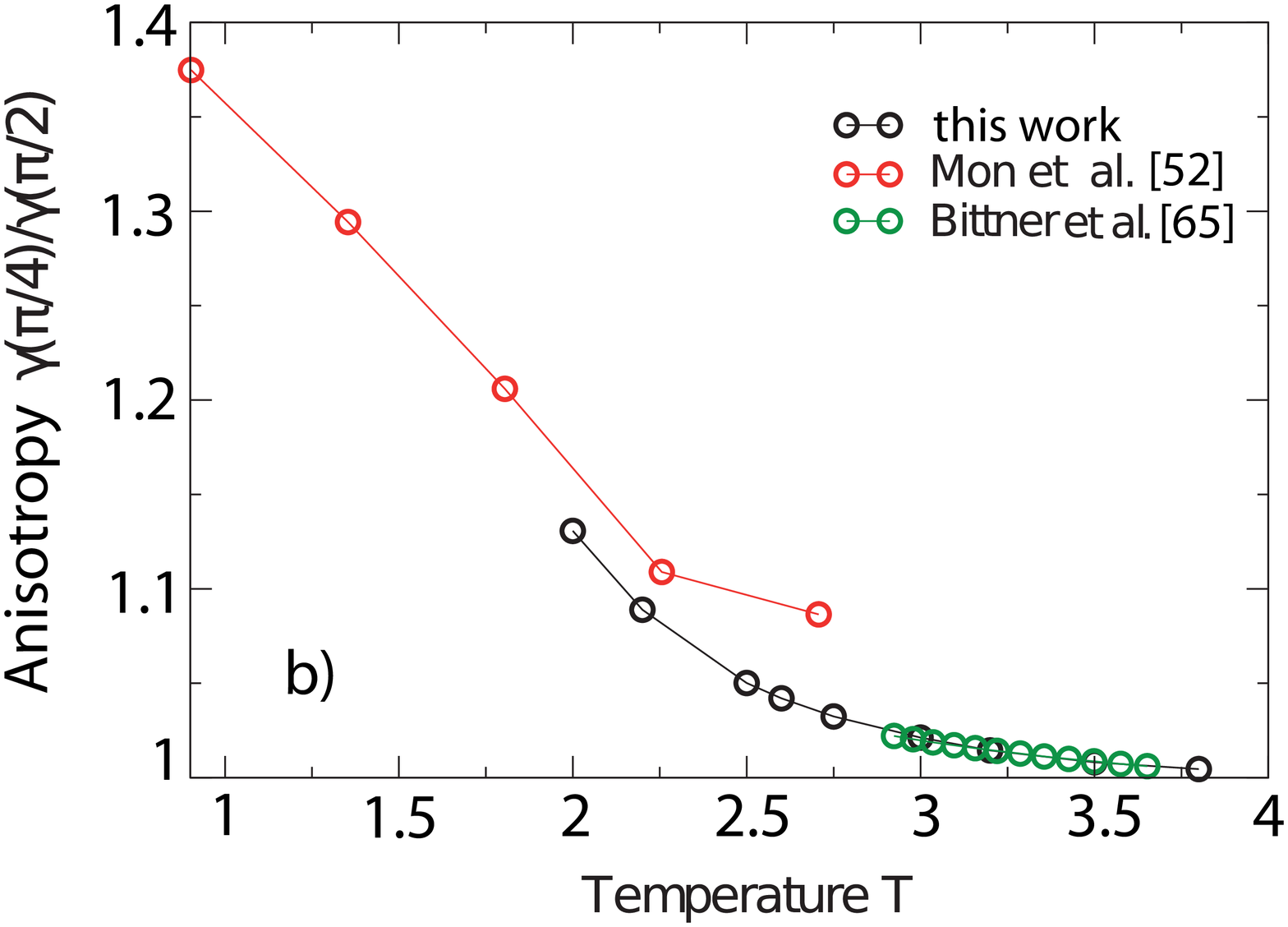}
\caption{\label{fig6} (Color online) Maximum surface tension anisotropy $\gamma(\pi/4)/\gamma(\pi/2)$ plotted versus reduced temperature $T$. Case (a) focuses on the regime $T > T_R$, comparing the present results with data due to Bittner et al. \cite{65}. Case (b) includes the low temperature region, adding also the estimates due to Mon et al. \cite{52} (uppermost curve).}
\end{figure}

\begin{figure}
\centering
\includegraphics [scale=0.28]{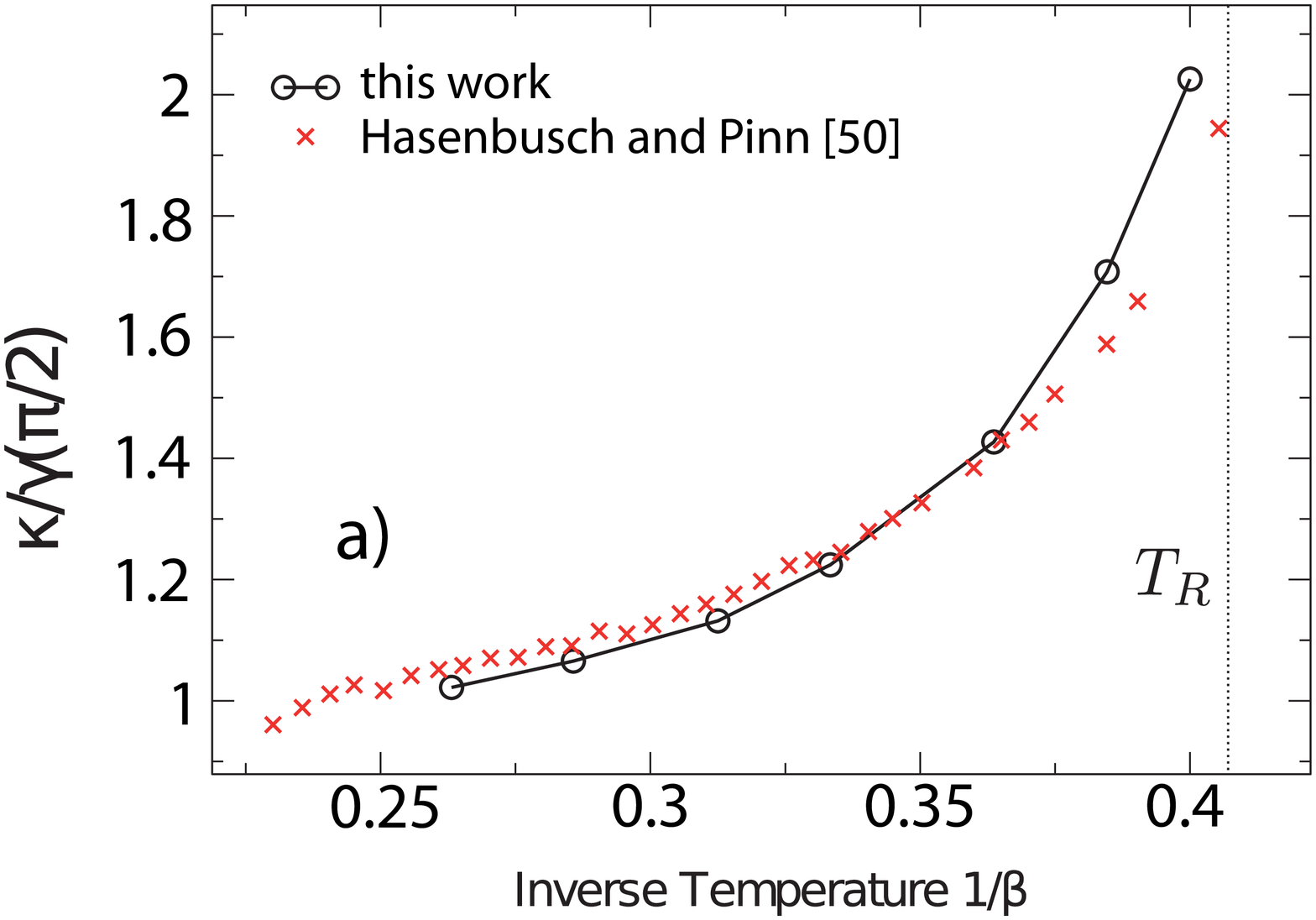}
\includegraphics [scale=0.28]{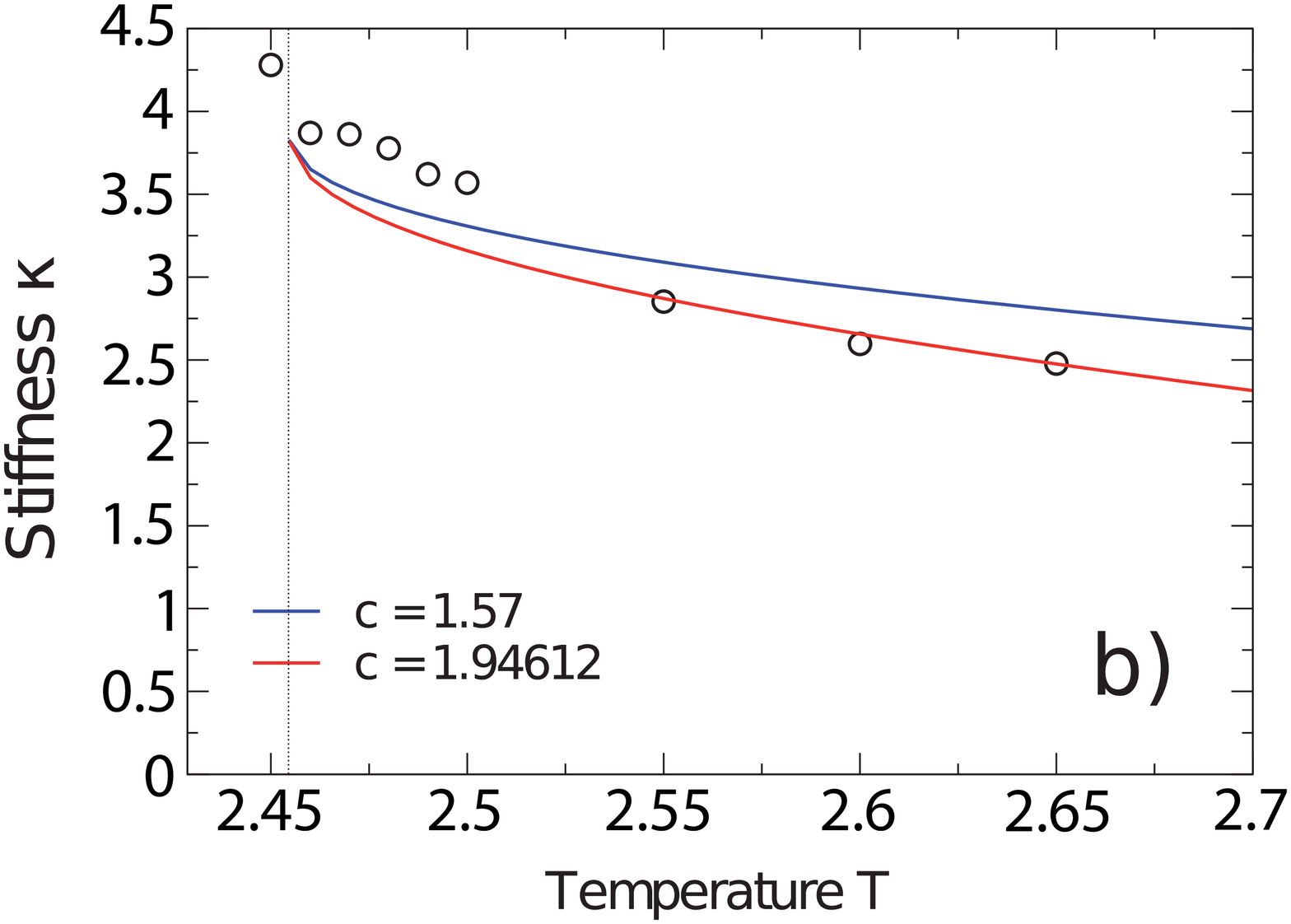}
\caption{\label{fig7} (Color online) (a) Interfacial stiffness $\kappa$ normalized by $\gamma (\pi/2)$ plotted vs. inverse temperature $\beta$, comparing the present estimation (circles) with the results from Hasenbusch and Pinn \cite{50} (crosses). Note that $\kappa/\gamma(\pi/2)$ must tend to unity as $T \rightarrow T_c$ (recall that $\beta _c \approx 0.2217$). Straight lines connecting our data are only drawn to guide the eye. Dotted vertical line indicates the location of the roughening transition. (b) Interfacial stiffness $\kappa$ plotted vs. temperature. Open circles are the present estimates, while the curves represent the theoretical formula, Eq.~\ref{eq25} with two different choices of the constant $c$, as indicated ($c=1.57$ for the upper curve and $c=1.94612$ for the lower curve). Again the dotted vertical line indicates the location of the roughening transition temperature. All data are for a system of size 184 $\times$ 504 $\times$ 504 lattices spacings.}
\end{figure}

\begin{figure}
\centering
\includegraphics [scale=0.37]{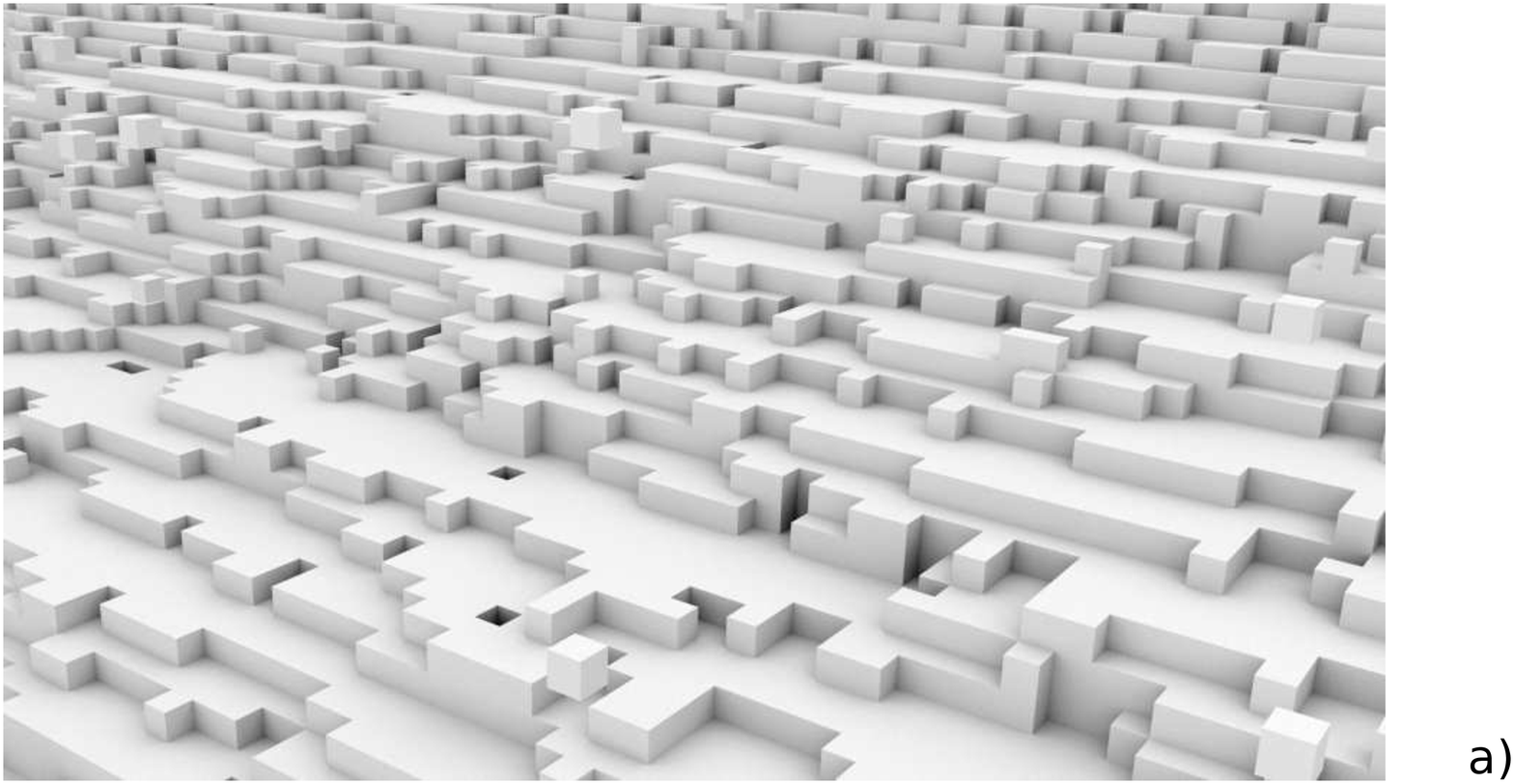}
\includegraphics [scale=0.37]{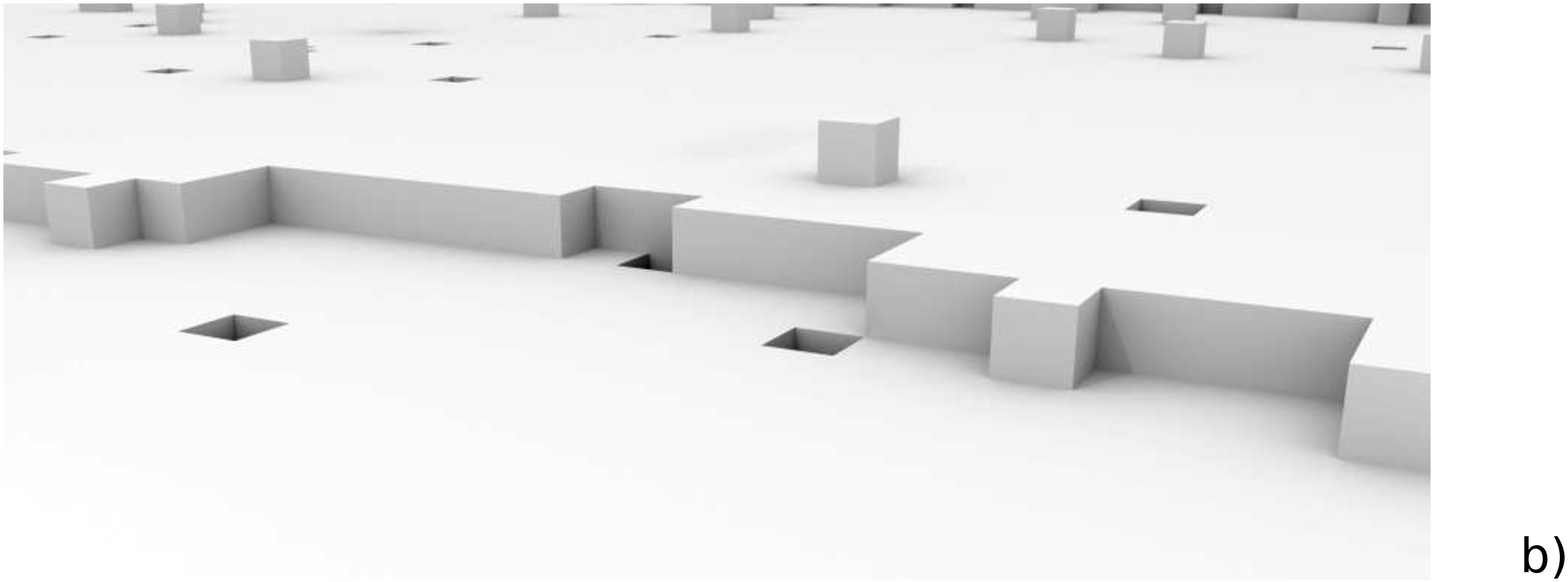}
\caption{\label{fig8} Snapshot picture of a tilted interface well below the roughening temperature (namely at $T=1.5$, for a system of size $184 \times 504 \times 504$), at wall fields $H_1=0.8$ (a) and 0.645 (b). Only small sections of the interface are displayed.}
\end{figure}

\begin{figure}
\centering
\includegraphics [scale=0.37]{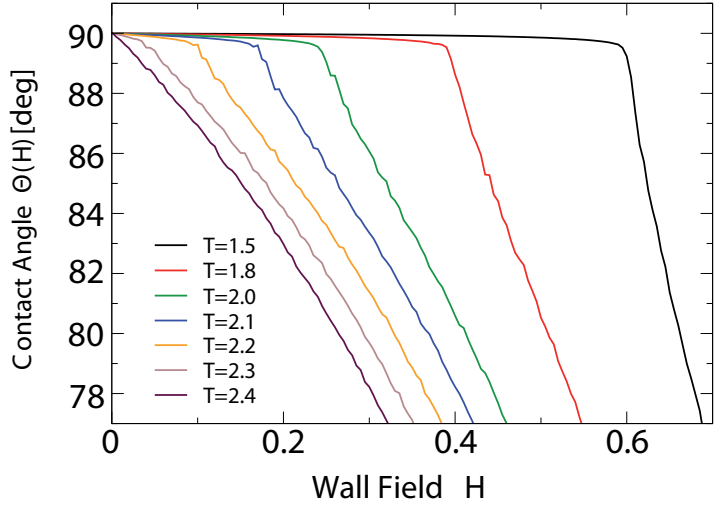}
\caption{\label{fig9} (Color online) Contact angle plotted versus the wall field $H_1$, for various temperatures in the range from $T=1.5$ (top) to $T=2.4$ (bottom), for systems with linear dimensions $L_x=184, \; L_y =L_z=504$.}
\end{figure}

\begin{figure}
\centering
\includegraphics [scale=0.37]{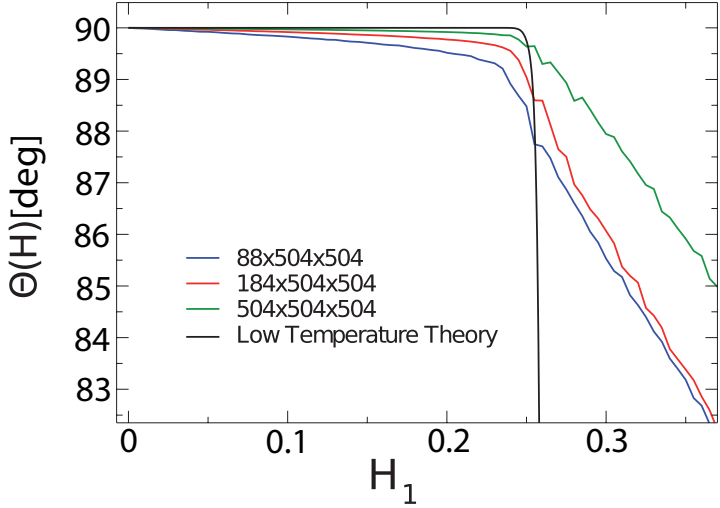}
\caption{\label{fig10} (Color online) Contact angle at $T=2.0$ plotted vs. $H_1$ for $L_y=L_z=504$ and three different choices of $L_x$ (from right to left: $L_x=504,184,88$). Estimating the step free energy at this temperature as $f_s(T)=0.52$, the low temperature approximation according to Eqs.~\ref{eq28} - \ref{eq31} is also shown (black curve).}
\end{figure}

\begin{figure}
\centering
\includegraphics [scale=0.37] {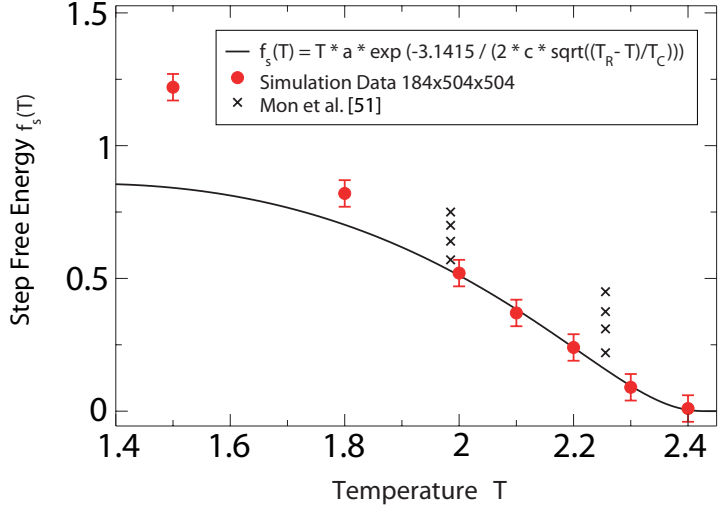}
\caption{\label{fig11} (Color online) Estimates for the step free energy plotted vs. temperature. Data points are estimates taken from the kink positions in Fig.~9, relying on the formula $f_s(T)\approx 2H_1^*$. The curve shows the formula $f_s(T)= const \exp \{-\pi/[2c\sqrt{(T_R-T/T_c)}]\}$, where the constant $c$ was fitted as $c=1.94 \pm 0.17$. Crosses denote estimates for $L_x=L_y=L_z=96,64,48$ and $32$ respectively (from bottom to top) from Mon et al. \cite{51}. 
}
\end{figure}

\begin{figure}
\centering
\includegraphics [scale=0.37]{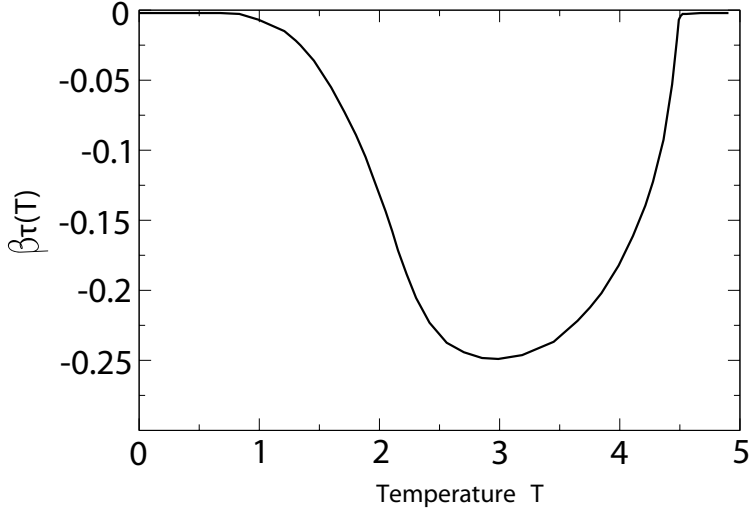}
\caption{\label{fig12} Temperature dependence of the line tension $\tau(0,\pi/2)$ (normalized by temperature $k_BT$) computed from the use of Eq.~\ref{eq20} using linear dimensions $L_z=L_x=20,30,40$ and 50, $L_y = 60$.}
\end{figure}

\begin{figure}
\centering
\includegraphics [scale=0.28]{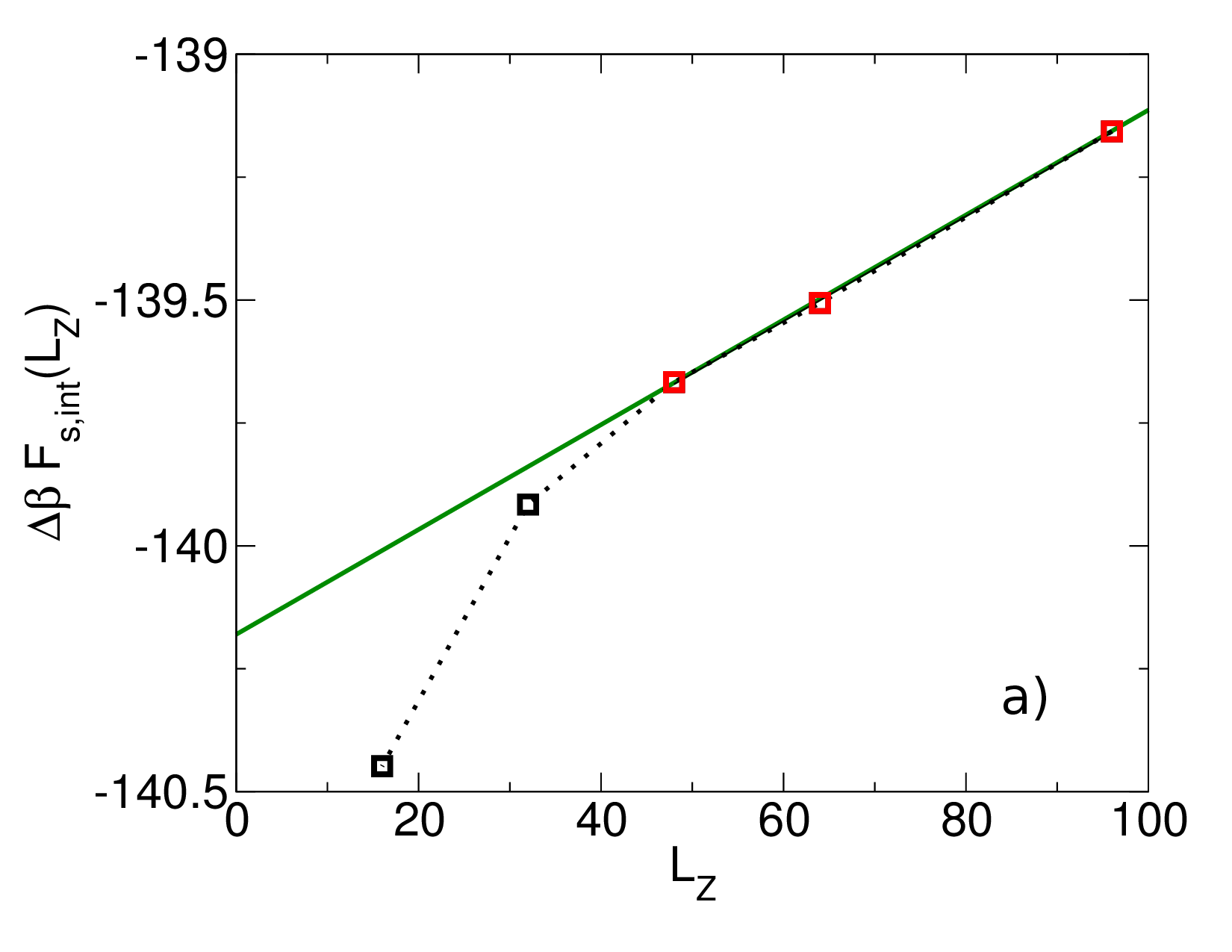}
\includegraphics [scale=0.28]{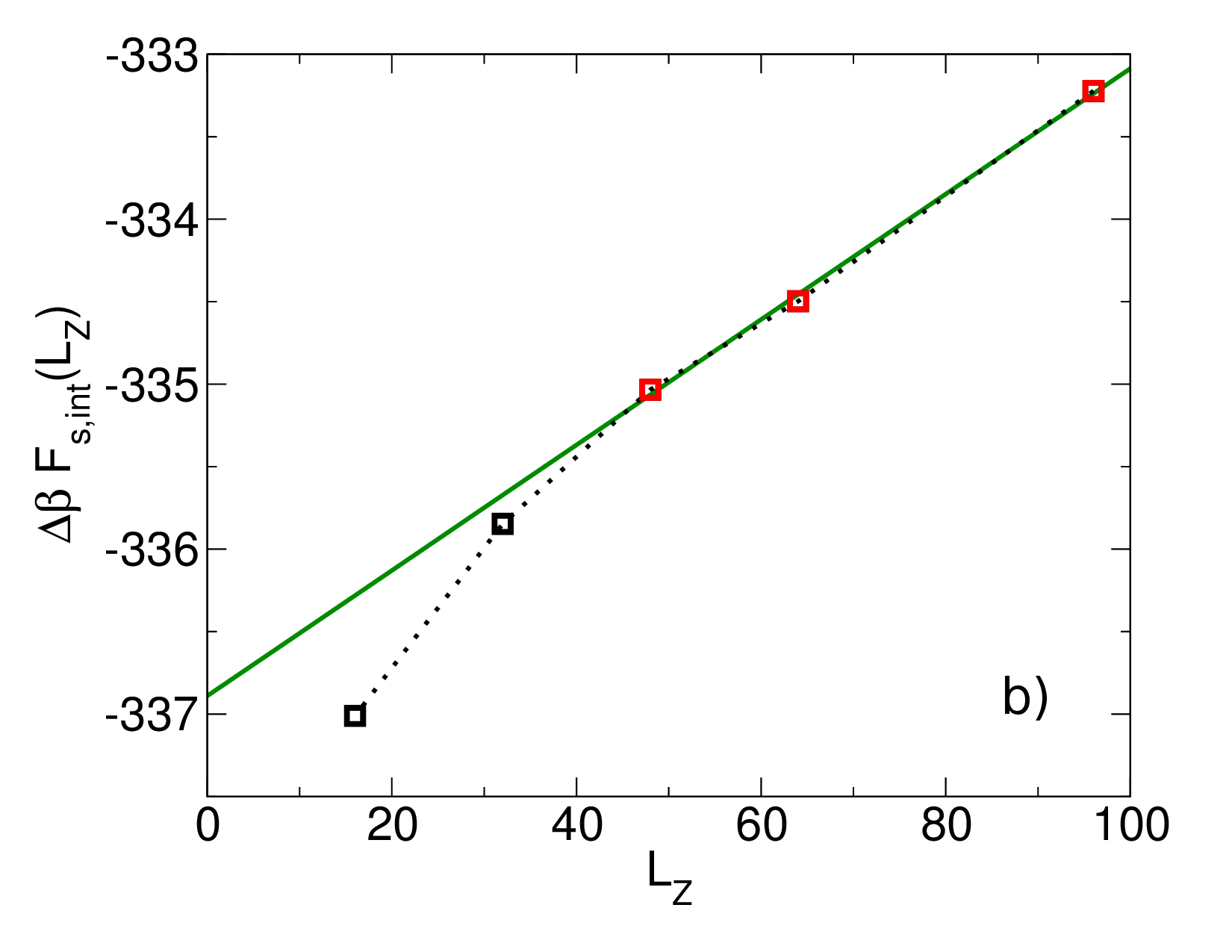}
\caption{\label{fig13} (Color online) Free energy differences $\Delta F (H_1)= - \int \limits _0 ^{H_1} [m_1(H_1')+m_n(H')]dH_1'$ \{Eq.~\ref{eq22}\} plotted vs. $L_z$ for $T=3.0$ and $H_1=0.35$ (a) and 0.45 (b), choosing $L_xL_z=1536$ and $L_yL_z = 9216$ constant.}
\end{figure}

\begin{figure}
\centering
\includegraphics [scale=0.37] {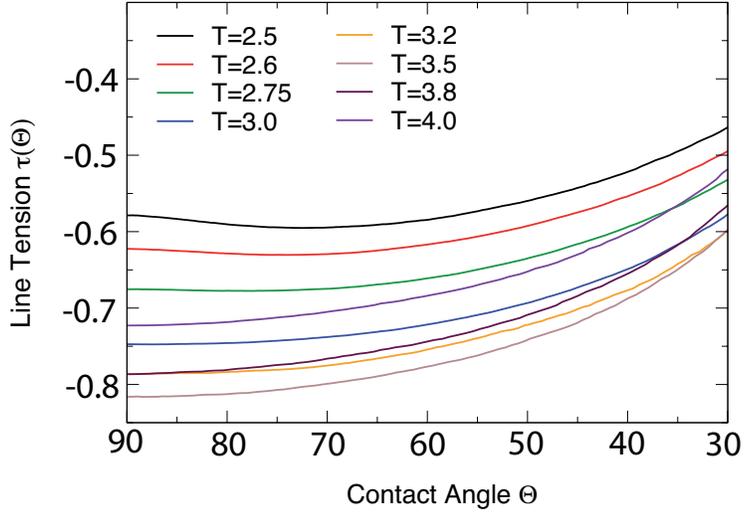}
\caption{\label{fig14} (Color online) Line tension $\tau(H_1,\theta)$ [in units of the exchange constant $J$] plotted as a function of contact angle $\theta$ for various temperatures, as indicated. Note that the most negative values of $\tau(H_1,\theta)$ occur for $T=3.5$. It should also be noted that finite-size effects occur for angles below 50$^{\circ}$.}
\end{figure}

\end{document}